\documentclass[12pt]{article}
\usepackage{amsmath}
\usepackage{graphicx}
\usepackage{enumerate}
\usepackage{natbib}
\usepackage{url} 
\usepackage{bm}
\usepackage{float} 
\usepackage{subfigure}
\usepackage{algorithm}
\usepackage{algpseudocode}
\usepackage{hyperref}
\usepackage{amsthm}
\usepackage{amsfonts}
\usepackage{amsmath, bm, amssymb, booktabs}
\usepackage{tikz}
\usetikzlibrary{shapes,arrows}
\usepackage{dsfont}
\usepackage{titling}
\usepackage{bbm}
\usepackage{enumitem}
\usepackage{diagbox}
\usepackage{comment}
\usepackage{pgfplots}
\usepackage{tikz}
\usetikzlibrary{shapes,backgrounds}
\usepackage{booktabs}
\usepackage{tabularx}
\usepackage{ltablex} 
\usepackage{makecell} 
\usepackage{tikz}
\usetikzlibrary{positioning}
\usepackage{setspace}
\usepackage{xr}
\usepackage{bbm}
\usepackage{authblk}
\usepackage{tikz}
\usetikzlibrary{shapes.geometric, arrows.meta, positioning}
\usepackage{booktabs} 
\usepackage{adjustbox} 
\usepackage{multirow} 
\usepackage{booktabs,siunitx,makecell,xcolor,threeparttable,tabularx}
\tikzstyle{startstop} = [rectangle, rounded corners, minimum width=3.5cm, minimum height=1cm,text centered, draw=black, fill=gray!20]
\tikzstyle{process} = [rectangle, minimum width=4cm, minimum height=1cm, text centered, draw=black, fill=blue!10]
\tikzstyle{decision} = [diamond, minimum width=4.5cm, minimum height=1.2cm, text centered, draw=black, fill=yellow!20]
\tikzstyle{arrow} = [thick,->,>=stealth]

\newcommand{\blind}{0}

\newtheorem{theorem}{\textbf Theorem}
\newtheorem{proposition}[theorem]{\textbf Proposition}

\DeclareMathOperator{\diag}{diag}

\begin{document}

\def\spacingset#1{\renewcommand{\baselinestretch}%
{#1}\small\normalsize} \spacingset{1}

\def\blind{1}
\if1\blind
{
  \title{\bf Generalized Heterogeneous Functional Model with Applications to Large-scale Mobile Health Data}
  \author[1]{Xiaojing Sun}
\author[2]{Bingxin Zhao}
\author[1]{Fei Xue}
\affil[1]{Department of Statistics, Purdue University}
\affil[2]{Department of Statistics and Data Science, University of Pennsylvania}
  \date{}
  \maketitle
} \fi

\if0\blind
{
  \bigskip
  \bigskip
  \bigskip
  \begin{center}
    {\LARGE\bf Title}
\end{center}
  \medskip
} \fi

\bigskip
\begin{abstract}
{
Physical activity is crucial for human health. With the increasing availability of large-scale mobile health data, strong associations have been found between physical activity and various diseases. However, accurately capturing this complex relationship is challenging,  possibly because it varies across different subgroups of subjects, especially in large-scale datasets.
To fill this gap, we propose a generalized heterogeneous functional method which simultaneously estimates functional effects and identifies subgroups within the generalized functional regression framework. The proposed method captures subgroup-specific functional relationships between physical activity and diseases, providing a more nuanced understanding of these associations. Additionally, we develop a pre-clustering method that enhances computational efficiency for large-scale data through a finer partition of subjects compared to true subgroups.
We further introduce a testing procedure to assess whether the different subgroups exhibit distinct functional effects.
In the real data application, we examine the impact of physical activity on the risk of dementia using the UK Biobank dataset, which includes over 96,433 participants. 
Our proposed method outperforms existing methods in future-day prediction accuracy, identifying three distinct subgroups, with detailed scientific interpretations for each subgroup. We also demonstrate the theoretical consistency of our methods. 
Codes implementing the proposed method are available at: \href{https://github.com/xiaojing777/GHFM}{https://github.com/xiaojing777/GHFM}.
}

\end{abstract}

\noindent%
\vfill

\newpage
\spacingset{1.9} 

{\color{black}
\section{Introduction}

{
Physical  activity plays a pivotal role in human health and is recognized as a factor that is highly related to the risk of both morbidity and mortality \citep{mahindru2023role, ekelund2019dose}. 
Modern advancements in technology, combined with the ubiquity of smartphones and wearable devices, have made access to physical activity data increasingly straightforward. 
In addition, large-scale mobile health datasets have become much more accessible over the past two decades \citep{althoff2017large}, such as the UK Biobank (UKB) dataset, which includes data from over 500,000 individuals \citep{sudlow2015uk}.
Many researchers study the relationship between physical activity and various diseases by analyzing large-scale datasets \citep{barker2019physical, rowlands2021association}. 
However, most of these studies focus on analyzing effects of physical activity on diseases that are assumed to be the same for all individuals, which might not be adequate for  large-scale datasets.

The heterogeneity in physical activity's effects across individuals could increase complexity 
of relationship between physical activity and diseases \citep{pearce2022association, rosmalen2012revealing}, and  this heterogeneity may arise from a range of complex factors. 
For example, differences in circadian rhythms may contribute to the effect heterogeneity:
the effect of physical activity on mortality for
individuals who go to bed early and get sufficient sleep 
could be different from that of
people who stay up late \citep{huang2022sleep}.
Exercise habits may also contribute to this variation: 
exercise beginners
often have more pronounced effects on health than people with long-established exercise routines \citep{garzon2017benefits}.
Although some studies  have investigated heterogeneity of physical activity \citep{albalak2023setting, shim2023wearable},
there are few studies considering heterogeneous effects of physical activity on diseases.


Moreover, existing studies focus mainly on summary statistics of physical activity, such as the minutes spent in moderate-to-vigorous physical activity \citep{rowlands2021association}.
In fact, physical activity is continuously varying across time, leading to a possibly time-dependent effect of physical activity on diseases. For instance, engaging in physical activity at midnight versus in the morning may lead to significantly different, or even opposite, effects on disease risk \citep{roig2016time}. 

To address this issue, 
functional data analysis (FDA) is a suitable method for capturing the time-dependent effects of physical activity \citep{ghosal2022scalar}. 
Some researchers have developed methods to account for heterogeneous functional effects \citep{zhou2025heterogeneous, yao2011functional, sun2022subgroup, zhang2022subgroup}. 
{
However, these methods are either computationally inefficient for large-scale mobile health datasets, not directly applicable to generalized outcome types, or require the number of subgroups to be specified in advance. 
To the best of our knowledge, there is no existing work that can simultaneously perform subgroup identification and coefficient function estimation under the scalar-on-function regression (SoFR) framework in a computationally feasible way for large-scale data and without requiring any pre-specified number of subgroups.

}

{
In this paper, we propose a generalized heterogeneous functional method  and a pre-clustering procedure under the SoFR framework for large-scale data to investigate the heterogeneous functional effects of physical activity on diseases. 
We propose to adopt subject-specific coefficient functions, rather than a single homogeneous coefficient function, to capture the varying effects of physical activity on diseases across individuals.  
We identify subgroups of subjects through 
fusing similar coefficient functions together based on a pairwise fusion penalty, where the coefficient function in each subgroup is assumed to be homogeneous.
Although subject-specific effects account for individual heterogeneity, they introduce a large number of parameters for large samples, particularly in the FDA framework, which results in high computational costs.

}


To address this, we propose a novel pre-clustering method for large-scale datasets, which reduces the number of parameters through 
grouping subjects into pre-clustering groups with most fitted coefficients, 
leading to a finer partition of true subgroups. 
{
Moreover, to validate whether the identified subgroups  exhibit heterogeneous functional effects, we also introduce a hypothesis testing procedure. The test compares subgroup-specific and homogeneous models, with the resulting statistic following a $\chi^2$ or $F$ distribution depending on the outcome type.
We apply the proposed methods to analyze the functional effects of physical activity on dementia using the UKB dataset.}

The novelty and  advantages of the proposed method are as follows. 
First, it effectively handles large mobile health datasets. 
 In contrast to homogeneous methods, our proposed method allows and can identify subgroups in large datasets to account for potential time-varying heterogeneous effects of physical activity on diseases without requiring a predetermined true number of subgroups, providing a more accurate
and comprehensive understanding of their underlying relationship. 
Second, our proposed pre-clustering technique reduces computation cost when dealing with large-scale data. Our theoretical analysis shows that
the proposed pre-clustering groups provide a finer partition of the true subgroup identification, and it does not break the true subgroup structure.
Third, our method can handle not only continuous responses, such as neuroticism scores, but also generalized outcomes from exponential family distributions, which is important for real-world applications (such as the binary dementia diagnosis outcome in the UKB study).

{
Fourth, in the real data application, our method outperforms existing approaches in terms of future-day prediction accuracy.
The proposed method identifies three subgroups in a dataset of 96,433 subjects with dementia diagnoses as outcomes.
Among the three subgroups for dementia, one has
a higher percentage of subjects diagnosed with dementia, facilitating the detection of dementia cases.
The analyses based on our identified subgroups are
consistent with 
existing studies \citep{ning2025accelerometer, iso2022physical, del2022association}. 
We also find that the optimal timing for physical activity to reduce disease risk varies across subgroups, providing practical recommendations for individualized interventions.
}

The remainder of this paper is organized as follows. 
In Section \ref{section: method}, we propose the generalized heterogeneous functional method, the testing procedure for heterogeneity of effects,  and the pre-clustering procedure. 
In section \ref{section: theory}, we illustrate theoretical results. Section \ref{section: simulation} provides numerical studies through simulations. In Section \ref{section: dementia application}, we apply the proposed method to the UKB dataset.  
}

{\color{black}
\section{Methodology}
\label{section: method}

{
In this section, we propose a generalized heterogeneous functional method (GHFM) to estimate  subgroup identification and coefficient functions simultaneously, and consider Gaussian and binary outcomes as specific examples. We further propose a pre-clustering method which improves computation efficiency for large sample while maintaining the true subgroup structure.
{
Moreover, we introduce a hypothesis testing procedure to validate whether the identified subgroups exhibit heterogeneous functional effects.}
}

\subsection{Generalized Heterogeneous Functional Method (GHFM)}

In this subsection, we develop a generalized subject-wise scalar-on-function model to
study inherent heterogeneity among subjects. 
Let $X_{i1}(t), \dots , X_{ip}(t)$ be $p$ functional covariates and $Y_i$ be a scalar response of the $i$-th subject for $t\in[0,T]$ and $i=1,\dots, n$.
Suppose that the $p$ functional covariates are observed at $m$ discrete time grids $t_1, t_2, \ldots, t_{m}$,
and that the response $Y_i$ follows an exponential family distribution with density 
$f_{Y_i}(y_i)=\exp \{(y_i \theta_i-b(\theta_i)) / a(\psi)-c(y_i, \psi)\}, i=1,2,\dots,n$, where $\theta_i$ is a canonical parameter, $\psi$ is a dispersion parameter, and $a(\cdot), b(\cdot), c(\cdot,\cdot)$ are known functions depending on specific distributions.

Here we allow  the parameter $\theta_i$ to be subject-specific to account for heterogeneity among subjects. Moreover, we propose a generalized heterogeneous functional regression model
\setlength\abovedisplayskip{3pt}
\setlength\belowdisplayskip{4pt}
\begin{equation}
    \label{Main Model of GHFM}
    \eta(\mu_i)    = \alpha + \sum_{j=1}^p \int_{0}^{T} X_{ij}(t)\beta_{ij}(t) dt \  \text{ for } \  i=1,2,\dots, n,
\end{equation}
where $\eta$ is a link function, $\mu_i = E(Y_i) = b'(\theta_i) $, $\alpha$ is the intercept term, and $\beta_{i1}(t), \dots, \beta_{ip}(t) $ are $p$ individualized  coefficient functions of the $i$-th subject defined on interval $[0,T]$. 
Rather than assuming a common coefficient function for each covariate in the traditional homogeneous scalar-on-function model, we consider subject-wise functional effects $\beta_{ij}(t)$ in the proposed model for $j=1,\dots, p$.

These subject-wise coefficient functions may exhibit certain subgroup structure of the entire population, which could be covariate-specific. 
Specifically, 
let $\mathcal{G}_j = \{\mathcal{G}_{j,1},\dots, \mathcal{G}_{j, \mathcal{K}_j} \}$ be a subgrouping partition of $\{1,\dots,n\}$
formed by coefficients $\beta_{1j}(t), \dots, \beta_{nj}(t)$ of the $j$-th covariate, where $\mathcal{K}_j$ is the number of distinct subgroups.
That is, for any $i, i' \in \{1,\dots,n\}$, we have $i, i' \in \mathcal{G}_{j,k}$ for some $k\in\{1,\dots, \mathcal{K}_j\}$ if and only if  $\beta_{ij}(t) = \beta_{i'j}(t)$.
Then the effect in each subgroup is homogeneous, while effects in different subgroups are distinct. This subgroup structure may vary across covariates.

To identify the subgroup structure and estimate  coefficient functions, we propose a pairwise grouping loss function
\begin{equation}
\label{loss func: general case}
\small
	Q_n(\alpha, \beta) = -\frac{1}{n}l(\alpha, \beta, \psi ; \mathbf{y}) + \phi \sum_{j=1}^p\sum_{i=1}^n  \int_0^{T} \left[\frac{d^2 \beta_{ij}(t)}{dt^2}\right]^2 dt + \lambda\sum_{j=1}^p \sum_{i\neq i'} \int_0^T |\beta_{ij}(t)-\beta_{i'j}(t)|dt,
\end{equation}
where 
{
\setlength\abovedisplayskip{2pt}
\setlength\belowdisplayskip{4pt}
$$l(\alpha, \beta, \psi ; \mathbf{y})=\sum_{i=1}^n l_i\left(\theta_i, \psi ; y_i\right)=\sum_{i=1}^n\left\{\left[y_i \theta_i(\alpha, \beta_i)-b\left(\theta_i(\alpha, \beta_i)\right)\right] / a(\psi)-c_i\left(y_i, \psi\right)\right\}$$
}
is the log likelihood function,
$\mathbf{y} = (Y_1,\dots,Y_n)^T, \beta_i = (\beta_{i1}(t), \dots, \beta_{ip}(t))^T$, $\beta = (\beta_{1}^T, \dots, \beta_{n}^T)^T$, and  $\phi, \lambda$ are tuning parameters.

Here we adopt a roughness penalty (the second term in the loss function) to control the fluctuation of each coefficient function $\beta_{ij}(t)$ \citep{green1993nonparametric}, and a functional pairwise fusion penalty (the third term in the loss function) \citep{ma2017concave} to fuse similar subject-wise effects together.
Specifically, the fusion penalty encourages subjects to share the same coefficient function when their corresponding effects are similar. 
In this way, we can reduce the number of coefficients, identify subgroups, and borrow information across subjects to estimate the coefficient functions accurately.

For estimation of the coefficient functions,
we propose to approximate $\beta_{ij}(t)$
through a linear combination of B-spline basis functions  for $i=1,\dots,n$ and $j=1,\dots,p$. 
Specifically, we approximate $\beta_{ij}(t) $ as ${\beta}_{ij}(t) \approx {\bm{b}}_{ij}^T \bm{B}(t)$,
where ${\bm{b}}_{ij}$ is an $L$-dimensional coefficient vector, 
$\mathbf{B}(t)$ is an $L$-dimensional vector of B-spline basis functions of order $(d+1)$ with $(M+1)$ equally spaced knots on the time interval $[0,T]$, and $L=M+d $. 
Given this approximation, the loss function in Equation (\ref{loss func: general case}) becomes
 \begin{equation}
 \small
 \setlength\abovedisplayskip{3pt}
\setlength\belowdisplayskip{2pt}
 	\label{loss func: general case b}
 	D_n(\alpha, \bm{b}_1, \dots, \bm{b}_p) = -\frac{1}{n}\tilde{l}(\alpha, \bm{b}_1, \dots, \bm{b}_p, \psi ; \mathbf{y}) + \phi\sum_{j=1}^p \bm{b}_j^TR\bm{b}_j + \lambda \sum_{j=1}^p  \sum_{i\neq i'} \int_0^T | \bm{B}^T(E_i - E_{i'})^T \bm{b}_j |dt,
 \end{equation}
 where $\bm{b}_j = ({\bm{b}}_{1j}^T,\dots, {\bm{b}}_{nj}^T)^T $ is an $(nL)$-dimensional column vector, $\tilde{l}(\alpha, \bm{b}_1, \dots, \bm{b}_p, \psi ; \mathbf{y})$ is the value of $l(\alpha, \beta, \psi ; \mathbf{y})$ with each $\beta_{ij}$ replaced by ${\bm{b}}_{ij}^T \bm{B}(t)$,
 \begin{equation}
\small
R = \diag\left\{\int _0^{T}  \frac{d^2 \mathbf{B}(t)}{dt^2} \frac{d^2 \mathbf{B}^T(t)}{dt^2} dt, \dots, \int _0^{T}  \frac{d^2 \mathbf{B}(t)}{dt^2} \frac{d^2 \mathbf{B}^T(t)}{dt^2} dt\right\}
\label{Matrix: R}
\end{equation}
is a block diagonal matrix with $n$ $L\times L$ blocks on the diagonal, and 
\begin{equation}
\setlength\abovedisplayskip{3pt}
\setlength\belowdisplayskip{4pt}
\label{def_E_i}
    E_i=(\bm{0}_{L\times L}, \dots, \bm{I}_{L\times L}, \dots, \bm{0}_{L\times L})^T
\end{equation}
is an $nL\times L$ matrix consisting of $n$  $L\times L$ blocks with the $i$-th block being an identity matrix of size $L$ and the other blocks being zero matrices.

We obtain the proposed estimator for coefficient functions by $\hat{\beta}_{ij}(t) = \hat{\bm{b}}_{ij}^T \bm{B}(t)$, where $\hat{\bm{b}}_{ij}$ is the $\bm{b}_{ij}$-coordinate of the minimizer of the loss function in Equation (\ref{loss func: general case b}).
Subject $i$ and Subject $i'$ are estimated to belong to the same subgroup for the $j$-th covariate if and only if $\hat{\beta}_{ij}(t) = \hat{\beta}_{i'j}(t)$ for $ i, i'=1,\dots,n$ and $j=1,\dots,p$.

{
The tuning parameters $\phi$ and $\lambda$ are selected based on the Bayesian Information Criterion (BIC), which in our context is defined as $\text{BIC} = k \ln(n) - 2 \ln(\hat{l})$, where $k$ is the total number of estimated parameters, and $\hat{l}$ is the likelihood evaluated at the estimated coefficient functions. More specifically, 
if there are $\hat{\mathcal{K}}_j$ estimated subgroups corresponding to the $j$-th covariate,
then $k = \sum_{j=1}^p (\hat{\mathcal{K}}_j L+1)$.
}

\subsection{Heterogeneous Functional Gaussian Method (HFGM)}

In this subsection, we consider the proposed method with Gaussian outcomes,  and refer to it as the heterogeneous functional Gaussian method (HFGM). Specifically, we assume $Y_i \sim  N(\mu_i, \sigma^2)$ for $ i=1,2,\dots,n$, {
implying that $\theta_i = \mu_i, b(\theta_i) = \mu_i^2/2, \eta(\mu_i) = \mu_i, a(\psi) = \sigma^2,$ and $c(y_i, \psi)=\left(y_i^2 / \sigma^2+\log 2 \pi \sigma^2\right) / 2$. 
}
For this Gaussian outcome, the proposed generalized heterogeneous functional regression model becomes
\begin{equation}
\setlength\abovedisplayskip{3pt}
\setlength\belowdisplayskip{4pt}
\label{model: HFGM}
Y_i = \alpha + \sum_{j=1}^p \int_{0}^{T} X_{ij}(t)\beta_{ij}(t) \, dt + \epsilon_i \ \text{ for } \  i=1,2,\dots,n,
\end{equation}
where $\epsilon_i \sim  N(0, \sigma^2).$ 
The corresponding Gaussian log likelihood is $l(\alpha, \beta, \psi ; \mathbf{y}) = \sum_{i=1}^n\{-(Y_i - \alpha - \sum_{j=1}^p \int_{0}^{T} X_{ij}(t)\beta_{ij}(t) dt)^2/2\sigma^2 - \log \left(2 \pi \sigma^2\right) / 2 \}$. 

Since maximizing $l(\alpha, \beta, \psi ; \mathbf{y})$ with respect to $\alpha$ and $\beta(t)$ is equivalent to minimizing the functional least squares term $ {l}^*(\alpha, \beta, \psi ; \mathbf{y}) = \sum_{i=1}^n(Y_i - \alpha - \sum_{j=1}^p \int_{0}^{T} X_{ij}(t)\beta_{ij}(t) dt)^2$, we propose the following loss function for the HFGM:
\begin{equation}\small
\setlength\abovedisplayskip{4pt}
\setlength\belowdisplayskip{4pt}
\label{loss func: gauss beta t}
Q_n^{G}(\alpha, \beta) = \frac{1}{n}{l}^*(\alpha, \beta, \psi ; \mathbf{y}) + \phi \sum_{j=1}^p \sum_{i=1}^n \int_0^{T} \left(\frac{d^2 \beta_{ij}(t)}{dt^2}\right)^2 dt + \lambda \sum_{j=1}^p \sum_{i\neq i'} \int_0^T |\beta_{ij}(t)-\beta_{i'j}(t)|dt.
\end{equation}
Using the B-spline basis functions, we can approximate this loss function by
{\small
\setlength\abovedisplayskip{3pt}
\setlength\belowdisplayskip{4pt}
\begin{align}
    D_n^G(\alpha, \bm{b}_1, \dots, \bm{b}_p) =& \frac{1}{n} \sum_{i=1}^n \left( Y_i - \alpha - \sum_{j=1}^p \bm{\gamma}_{ij}^T E_i^T \bm{b}_j  \right)^2 + \phi \sum_{j=1}^p \bm{b}_j^TR\bm{b}_j \notag\\
    &+ \lambda \sum_{j=1}^p \sum_{i\neq i'} \int_0^T | \bm{B}^T(E_i - E_{i'})^T \bm{b}_j |dt, \label{loss func: gaussian b}
\end{align}
}
where $\bm{\gamma}_{ij} = \int_0^T \bm{B}(t)X_{ij}(t)dt $ is an $L$-dimensional column vector. 
 
\subsection{Heterogeneous Functional Logistic Method (HFLM)}

In this subsection, we consider the proposed method with binary outcomes which are common in practical scenarios, particularly in medical health where diagnosed cases are denoted as "1" and non-diagnosed cases as "0". 
We refer to the proposed method for binary outcomes as the heterogeneous functional logistic  method (HFLM). 
Here we assume that the binary response $Y_i$ follows the Bernoulli distribution with probability $p_i$ for $i=1,2,\dots, n$. Then the heterogeneous functional logistic regression model is of the form
\begin{equation}
\setlength\abovedisplayskip{3pt}
\setlength\belowdisplayskip{4pt}
\label{model: HFLM}
    \ln [{p_i}/{(1-p_i)}] = \alpha + \sum_{j=1}^p \int_{0}^{T} X_{ij}(t)\beta_{ij}(t) dt,  i=1,\dots, n.
\end{equation}
The corresponding loss function for the HFLM is
\begin{equation}
\small
\setlength\abovedisplayskip{3pt}
\setlength\belowdisplayskip{4pt}
\label{loss func: logistic beta t}
Q_n^L(\alpha, \beta) = -\frac{1}{n}l(\alpha, \beta, \psi ; \mathbf{y}) + \phi \sum_{j=1}^p \sum_{i=1}^n  \int_0^{T} \left[\frac{d^2 \beta_{ij}(t)}{dt^2}\right]^2 dt + \lambda \sum_{j=1}^p \sum_{i\neq i'} \int_0^T |\beta_{ij}(t)-\beta_{i'j}(t)|dt,
\end{equation}
where
{\small
\begin{equation*}
	\setlength\abovedisplayskip{1pt}
	\setlength\belowdisplayskip{4pt}
	l(\alpha, \beta, \psi ; \mathbf{y}) = \sum_{i=1}^n \left\{ y_i\left[\alpha+ \sum_{j=1}^p\int\beta_{ij}(t)X_{ij}(t)dt\right] - \ln\left[1+\exp\left(\alpha+\sum_{j=1}^p\int\beta_{ij}(t)X_{ij}(t)dt\right) \right]  \right\}.
\end{equation*}}

\noindent Using the B-spline basis functions, we approximate this loss function by
\begin{equation}
\small
\setlength\abovedisplayskip{3pt}
\setlength\belowdisplayskip{4pt}
\label{loss func b: logis}
D_n^L(\alpha,\bm{b}_1, \dots, \bm{b}_p) = -\frac{1}{n}\tilde{l}(\alpha, \bm{b}_1, \dots, \bm{b}_p, \psi ; \mathbf{y}) + \phi \sum_{j=1}^p \bm{b}_j^TR\bm{b}_j + \lambda \sum_{j=1}^p  \sum_{i\neq i'} \int_0^T | \bm{B}^T(E_i - E_{i'})^T \bm{b}_j |dt,
\end{equation}
where
$\tilde{l}(\alpha, \bm{b}_1, \dots, \bm{b}_p, \psi ; \mathbf{y}) = \sum_{i=1}^n [ y_i (\alpha+ \sum_{j=1}^p \gamma _{ij}^TE_i^T\bm{b}_j) - \ln(1+\exp{(\alpha + \sum_{j=1}^p \gamma _{ij}^T E_i^T\bm{b}_j)} )] $. 

\subsection{Pre-clustering}

{
Since parameters in the proposed loss functions in Equations (\ref{loss func: general case}) and (\ref{loss func: general case b}) are subject-wise, the minimization of the proposed loss could be computationally challenging when the sample size is large. In fact, this kind of large-sample or large-scale data are common in practice.
For instance, the number of subjects in the UKB study with available physical activity data is  93,670. }


To solve this issue, we propose a pre-clustering method, which provides an initial clustering of all subjects.
A pre-clustering is a partition of all subjects into
$K$ potential groups such that, 
if the $i$-th subject and the $i'$-th subject are in the same potential group, then we have $\beta_{ij}(t) = \beta_{i'j}(t)$ for all $j=1,\dots, p$. That is, subjects in the same potential group belong to the same true subgroup for each covariate. 
We refer to these potential groups as pre-clustering groups. 
Note that
$K\ge\prod_{j=1}^p\mathcal{K}_j$.

Subjects in the same true subgroup might not be in the same pre-clustering group, which implies that the pre-clustering group structure is a finer partition of all subjects than the true subgroup structure.
Figure \ref{fig:Relationship between true subgroups and pre-clustering groups. } illustrates this through an example with two true subgroups and five pre-clustering groups, where the first true subgroup consists of three pre-clustering groups, and the second true subgroup consists of two pre-clustering groups.
Moreover, the pre-clustering group structure is not covariate-specific, that is, there is only one rather than $p$ pre-clustering group structures.

\begin{figure}[h]
	\centering
	\includegraphics[width=0.65\textwidth]{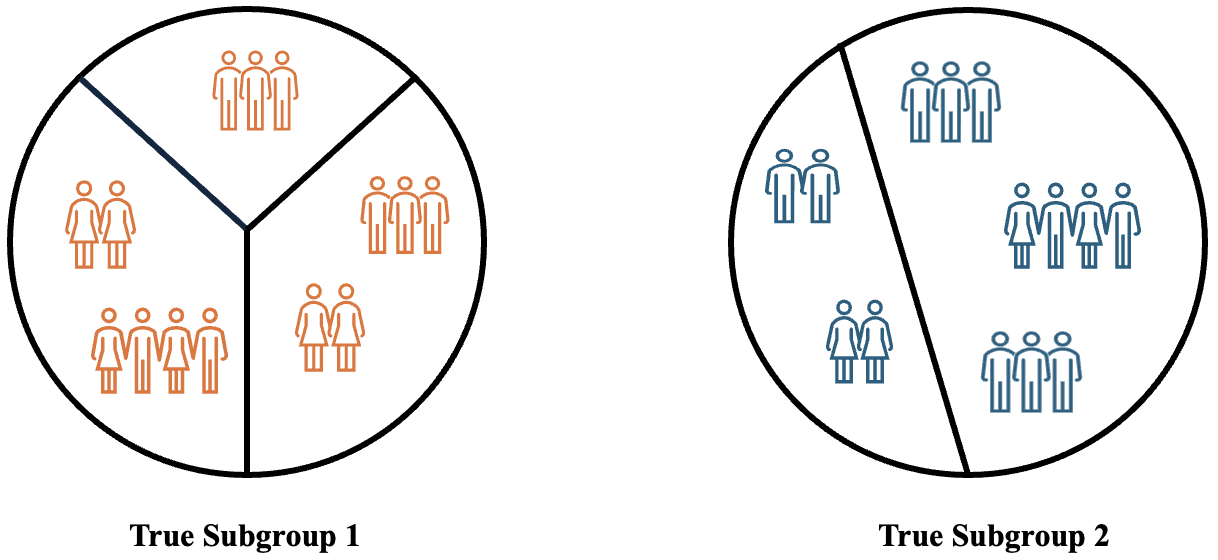}
	\caption{An example for relationship between true subgroups and pre-clustering groups when there are two true subgroups and five pre-clustering groups.}
	\label{fig:Relationship between true subgroups and pre-clustering groups. }
\end{figure}

{

We propose a pre-clustering loss function
\begin{equation}
\setlength\abovedisplayskip{1pt}
\setlength\belowdisplayskip{4pt}
\small
    L(\alpha, \beta_{(1)},\dots,\beta_{(K)}) = \sum_{i=1}^{n} \min_{k= 1,\dots,K} L_{i,k}(\alpha, \beta_{(k)}),
\label{loss function: pre-clustering}
\end{equation}
where $\alpha$ is the intercept term, $\beta_{(k)} = (\beta_{(k),1}(t), \dots, \beta_{(k),p}(t))^T$ with $\beta_{(k),j}(t)$ denoting the coefficient function  in the $k$-th pre-clustering group for the $j$-th covariate, and $L_{i,k}(\alpha, \beta_{(k)}) = -\left[y_i \theta_i(\alpha, \beta_{(k)})-b\left(\theta_i(\alpha, \beta_{(k)})\right)\right] / a(\psi)+c_i\left(y_i, \psi\right)$ is the negative log-likelihood of the $i$-th subject with parameters $\alpha$ and $\beta_{(k)}$ from the $k$-th pre-clustering group. 
Let 
\begin{equation}
\setlength\abovedisplayskip{3pt}
\setlength\belowdisplayskip{4pt}
\label{minimizer: pre-clustering}
(\tilde{\alpha}, \tilde{\beta}_{(1)},\dots,\tilde{\beta}_{(K)})  = 
\underset{\alpha \in \mathbb{R}, {\beta}_{(k),j}(t) \in S_{dM}, k\in\{1,\dots,K\}, j\in\{1,\dots,p\} }{\mathrm{argmin}}
L({\alpha}, {\beta}_{(1)},\dots,{\beta}_{(K)}),
\end{equation}
be the minimizer of the pre-clustering loss function, where $\mathcal{S}_{d M}$ denotes the space spanned by B-spline basis functions of degree $d+1$ and knots $M+1$.

Here we identify the pre-clustering group label of the $i$-th subject through finding out the smallest $L_{i,k}$ among all pre-clustering groups ($k=1,\dots, K$) since a smaller loss indicates better fitness between the subject and a pre-clustering group.
Following this rule, we can obtain pre-clustering groups $\tilde{\mathcal{G} }= \{ \tilde{\mathcal{G}}_{1},\dots, \tilde{\mathcal{G}}_{K} \}$
identified by the minimizer $(\tilde{\alpha}, \tilde{\beta}_{(1)},\dots,\tilde{\beta}_{(K)})$, 
{where $\tilde{\mathcal{G}}_{k}$ is the index set of subjects in the $k$-th pre-clustering group for $k=1,\dots, K$.}

Compared with the loss function in Equation (\ref{loss func: general case}), the above pre-clustering loss function reduces the number of coefficient functions from $n$ to $K$ for each covariate since the coefficient functions in Equation (\ref{loss function: pre-clustering}) are group-specific instead of subject-specific, which reduces computation cost.
In Theorem \ref{thm: pre-clustering} of Section \ref{section: theory} below, we will show that minimizing  the pre-clustering loss can lead to an estimated pre-clustering
group identification that does not break the true subgroup structure 
under some regularity conditions. 
}

After the pre-clustering groups are estimated, we apply the proposed pairwise grouping loss function in Equation (\ref{loss func: general case}) to the identified pre-clustering groups, and 
minimize
the following loss function
\begin{equation}
\setlength\abovedisplayskip{3pt}
\setlength\belowdisplayskip{4pt}
\begin{split}
\label{loss func: after pre-clustering}
\small
Q_n(&\alpha,  \beta_{(1)},\dots,\beta_{(K)} ) = -\Bar{l}(\alpha,  \beta_{(1)},\dots,\beta_{(K)}, \psi ; \mathbf{y}) \\
&\quad + \phi \sum_{j=1}^p \sum_{k=1}^K  \int_0^{T} \left(\frac{d^2 \beta_{(k),j}(t)}{dt^2}\right)^2 dt  + \lambda \sum_{j=1}^p \sum_{k\neq k'} \int_0^T |\beta_{(k),j}(t)-\beta_{(k'),j}(t)|dt,
\end{split}
\end{equation}
where 
$\Bar{l}(\alpha,  \beta_{(1)},\dots,\beta_{(K)}, \psi ; \mathbf{y}) = \sum_{k=1}^K  \sum _{i\in \tilde{\mathcal{G}}_{k}} \left\{\left[y_i \theta_i(\alpha, \beta_{(k)})-b\left(\theta_i(\alpha, \beta_{(k)})\right)\right] / a(\psi)-c_i\left(y_i, \psi\right)\right\} $.
{
Subjects in the $k$-th pre-clustering group and in the $k'$-th pre-clustering group are estimated to belong to the same subgroup for the $j$-th covariate if and only if $\hat{\beta}_{(k),j}(t) = \hat{\beta}_{(k'),j}(t)$, where $\hat{\beta}_{(k),j}(t)$ is the $\beta_{(k),j}(t)$-coordinate of the minimizer of the loss in Equation (\ref{loss func: after pre-clustering}) for $ k, k'=1,\dots,K$ and $j=1,\dots,p$.
Consequently, $\hat{\beta}_{(k),j}(t)$ is the estimated coefficient function in this subgroup.
}

\begin{figure}[H]
\centering
\includegraphics[width=0.8\textwidth,height=0.37\textwidth]{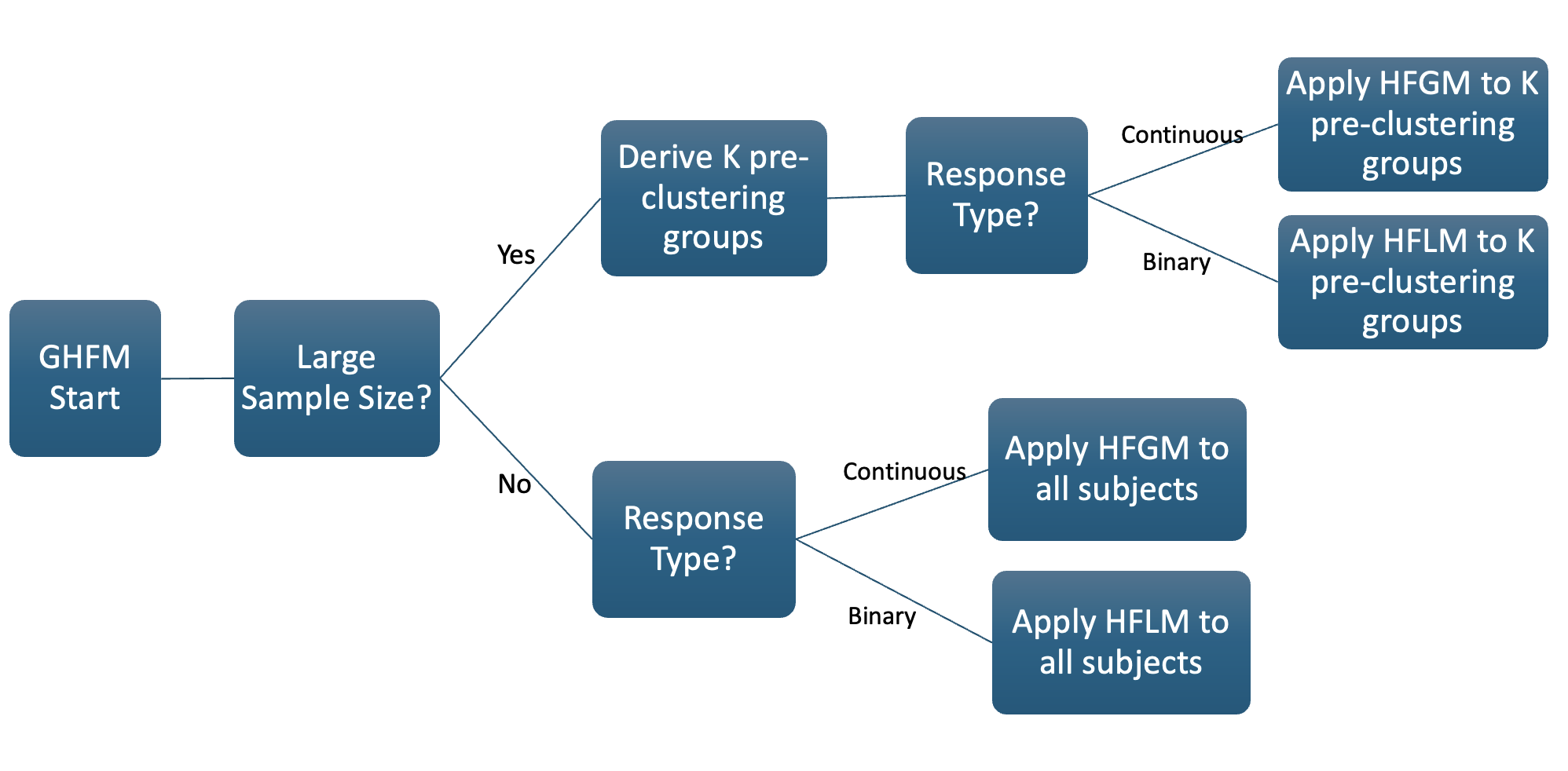}
\caption{The flowchart of the proposed  method.}
\label{fig:Flowchart}
\end{figure}

{
To conclude this subsection, we summarize the overall workflow of the proposed method in Figure~\ref{fig:Flowchart}. 
In practice, we treat datasets with more than 500 subjects as ``large sample" cases, since running GHFM on all individuals becomes computationally intensive and often exceeds memory limits based on simulations in Section \ref{section: simulation}. 
As a remedy, we introduce a pre-clustering step to partition the data before executing GHFM.
Accordingly, the number of pre-clustering groups \( K \) needs to be chosen with this computational constraint in mind. 
We therefore recommend using \( K \leq 500 \), with practical choices such as \( K = 50 \), \( 100 \), or \( 200 \) offering a balance between resolution and efficiency.
}

{
We provide detailed algorithmic implementations of the proposed GHFM and the pre-clustering method in  Supplementary Materials. For GHFM, we present two computational strategies: one based on linear quadratic approximation (LQA) \citep{fan2001variable}, and the other based on the alternating direction method of multipliers (ADMM) \citep{boyd2011distributed}.
}

{
\subsection{Generalized Test for Functional Subgroup Effects}

In this subsection, we propose a generalized likelihood ratio test for exponential family responses to assess subgroup heterogeneity. 
Specifically, given 
estimated subgroups for all covariates, 
we compare a full model \( \mathcal{M}_F \) that allows subgroup-specific functional coefficients with a reduced model \( \mathcal{M}_R \) that assumes a common effect across all subgroups. 
The null and alternative hypotheses are
{\small
\begin{equation}
\small 
\setlength\abovedisplayskip{3pt}
\setlength\belowdisplayskip{4pt}
\begin{aligned}
H_0:&\ \text{Functional effects are the same across all subgroups for all covariates}, \\
\text{vs. } \quad 
H_1:&\ \text{
At least two subgroups have different functional effects}.
\end{aligned}
\label{eq: H0 and Ha of testing effects in different groups}
\end{equation}
}
We construct the following test statistic for 
the hypotheses
in (\ref{eq: H0 and Ha of testing effects in different groups}) 
based on the deviance difference between the reduced and full models:
\begin{equation}
\small 
\setlength\abovedisplayskip{3pt}
\setlength\belowdisplayskip{4pt}
   \frac{D(\hat{\mu}^R; \mathbf{y}) - D(\hat{\mu}^F; \mathbf{y})}{\hat{\phi}}, 
   \label{eq: general test stat}
\end{equation}
where \( \hat{\mu}^R \) and \( \hat{\mu}^F \) are the estimated conditional means under the reduced (homogeneous) and full (subgroup-specific) models, respectively. 
Here \( D(\cdot; \cdot) \) denotes the model deviance and \( \hat{\phi} \) is the estimated dispersion parameter.

\vspace{1em}
\noindent \textbf{Example (HFGM):} For the proposed HFGM method, 
the test statistic in (\ref{eq: general test stat}) is given by
\[
\small 
\setlength\abovedisplayskip{3pt}
\setlength\belowdisplayskip{4pt}
F = \frac{(RSS_R - RSS_F)/\left((\sum_{j=1}^p\hat{\mathcal{K}}_j - p)L\right)}{RSS_F/\left(n - (\sum_{j=1}^p \hat{\mathcal{K}}_j )L - p \right)},
\]
where \( RSS_R \) and \( RSS_F \) are the residual sum of squares from the reduced and full models, respectively. 
Under the null hypothesis stated in (\ref{eq: H0 and Ha of testing effects in different groups}), this statistic follows an \( F \)-distribution with $(\sum_{j=1}^p\hat{\mathcal{K}}_j - p)L$ and \(n - (\sum_{j=1}^p \hat{\mathcal{K}}_j )L - p\) degrees of freedom.

\vspace{1em}
\noindent \textbf{Example (HFLM):} For the proposed HFLM method, 
the test statistic in (\ref{eq: general test stat}) becomes
\[
\setlength\abovedisplayskip{7pt}
\setlength\belowdisplayskip{8pt}
\small
\Lambda = D(\hat{\mu}^R; \mathbf{y}) - D(\hat{\mu}^F; \mathbf{y}),
\]
which approximately follows a $\chi^2_{(\sum_{j=1}^p\hat{\mathcal{K}}_j - p)L}$ distribution under $H_0$ in  (\ref{eq: H0 and Ha of testing effects in different groups}).

In practice, we apply the test to GHFM-identified subgroups to assess whether they capture heterogeneous effects.

}

\section{Theory}
\label{section: theory}


\subsection{Consistency of GHFM}
In this subsection, 
we  demonstrate consistency of subgroup identification and parameter estimation of the proposed GHFM.

We  first introduce some notations and 
regularity conditions.
We let $\lambda_n=\lambda$ and  $\phi_n=\phi$ in this section to emphasize that the tuning parameters may change as sample size $n$ increases. 
We define the $L_2$ norm and infinity norm of a continuous function $h$ on the interval $[0, T]$ as 
$
\|h(t)\|_2 = \left(\int_0^T h^2(t) \, dt\right)^{1/2}
$
and 
$\|h (t)\|_{\infty}=\sup \{|h(t)|: t \in[0, T]\}$
respectively.
We use  $f(n) \gtrsim g(n)$ to denote that there exists a constant $C > 0$ such that $f(n) \geq C \cdot g(n)$ for sufficiently large $n$. 
We let $\lambda_{\min}(A)$ and $\lambda_{\max}(A)$ represent the minimum and maximum eigenvalues of matrix $A$, 
and $J_{n,j} = \diag\{\bm{\gamma}_{1j}\bm{\gamma}_{1j}^T, \dots, \bm{\gamma}_{nj}\bm{\gamma}_{nj}^T\}$.



(C1) We assume that $\|X_j(t)\|_2$ is almost surely  bounded above by a constant for $j=1,\dots,p$, and that $X_{1j}(t),\dots,X_{nj}(t)$ are $n$  independent and identically distributed (i.i.d.) realizations of the stochastic process $X_j(t)$.



(C2) Suppose that, for each $i\in \{1,\dots,n\}$ and $j\in \{1,\dots,p\}$, the coefficient function $\beta_{ij}(t)$ is in the Hölder space $C^{p^*, v}$ for some positive integer $p^*$ and $v \in[0,1]$ such that $p^*+v\le d$ .
That is, 
there is a constant $C_1>0$ such that
$\mid \beta_{ij}^{\left(p^*\right)}\left(u_1\right)-\beta_{ij}^{\left(p^*\right)}\left(u_2\right)\left|\leq C_1\right| u_1-\left.u_2\right|^v$ for any
$u_1, u_2 \in [0,T]$, where $\beta_{ij}^{\left(p^*\right)}\left(t\right)$ denotes the $p^*$-th derivative of $\beta_{ij}(t)$.


(C3) 
We assume that the number of time grids $M=o\left(n^{1 / \psi_1 }\right)$, $M \gtrsim n^{1 / \psi_2 }$, 
tuning parameters 
$\phi_n=o\left(n^{-1 / 2}\right)$ and $\lambda_n=o(n^{-3/2} M^{-1})$, where $\psi_2>\psi_1>2$.

{
(C4) 
We assume $\max \{\bm{\gamma}_{1j}^T\bm{\gamma}_{1j}, \dots, \bm{\gamma}_{nj}^T\bm{\gamma}_{nj}\} \gtrsim M^{-1/2}$ almost surely for $j\in\{1,\dots,p\}$.

}


(C5) We assume the  number of true subgroups 
$\mathcal{K}_j = O(1)$ and $\min_{k=1,\dots,\mathcal{K}_j} | \mathcal{G}_{j,k} | \gtrsim n$  for $j\in\{1,\dots,p\}$.

(C6) 
Suppose that $\min _{i \in \mathcal{G}_{j,k}, i' \in \mathcal{G}_{j, k'}, k \neq k^{\prime}}\left\|\beta_{ij}(t)-\beta_{i'j}(t)\right\|_2 > C_3 
$ for some constant $C_3 > 0$.

 Conditions (C1) and (C2) are the same as Hypothesis (H1) and (H3) of \cite{cardot2003spline} which are standard assumptions for non-parametric B-spline methods \citep{shen1998local, xue2010consistent, claeskens2009asymptotic}. Condition
(C1) ensures the identifiability of penalized functional linear model, that is, the existence
and uniqueness of the coefficient functions \citep{bosq2000linear}.
Condition (C3) specifies the exact requirements for the tuning parameters.
Condition (C4) is equivalent to requiring that $\lambda_{\max}(J_{n,j}) \gtrsim M^{-1/2}$, which is similar to Condition (C4) in \cite{lin2017locally} and Condition (A8) in \cite{zhou2013functional}.}
{\color{black}
Condition (C5) indicates that the number of true subgroups does not increase as the sample size goes to infinity, while the size of each subgroup expands with increasing samples. 
Condition (C6)  provides a lower bound of the minimum distance of coefficient functions in different subgroups,
which is crucial for identifying heterogeneous subgroups \citep{ma2017concave}.


The following Theorem \ref{thm: consistency of HFGM} shows that with probability tending to one, there exists a local solution of Equation (\ref{loss func: gauss beta t}) that converges to $\beta_{1j}(t), \ldots, \beta_{nj}(t)$, and the corresponding estimated subgroup identification $\{\hat{\mathcal{G}}_{j,1},\dots, \hat{\mathcal{G}}_{j, \hat{\mathcal{K}}_j} \}$ converges to true subgroup identification $\{\mathcal{G}_{j,1},\dots, \mathcal{G}_{j, \mathcal{K}_j} \}$ for $j\in \{1,\dots,p\}$.

\begin{theorem}
\label{thm: consistency of HFGM}
When Conditions (C1) to (C6) are satisfied, for the loss function of the HFGM in Equation (\ref{loss func: gauss beta t}),
with probability tending to one, there exists a local minimizer  $(\hat{\alpha}, \hat{{\beta}}) = \arg \min_{\alpha\in\mathbb{R}, \beta \in \mathcal{S}_{d M}} Q_{n}^G({\alpha, \beta})$  such that 
$ \| \hat{\beta}_{ij}(t) - \beta_{ij}(t) \|_\infty < \xi_n$, where $\{\xi_n\}$ is a sequence converges to 0,
and  
$\operatorname{Pr}\left( \{\hat{\mathcal{G}}_{j,1},\dots, \hat{\mathcal{G}}_{j, \hat{\mathcal{K}}_j} \} =  \{\mathcal{G}_{j,1},\dots, \mathcal{G}_{j, \mathcal{K}_j} \} \right) \rightarrow 1$ for $i\in\{1,\dots,n\}, j\in \{1,\dots,p\}$,
where $ \hat{{\beta}} = (\hat{\beta}_{11}(t),\dots,\hat{\beta}_{np}(t))^T$.
\end{theorem}

The theorem demonstrates the consistency of both coefficient function estimation and subgroup identification. 
In fact, the consistency of subgroup identification naturally follows from the consistency of coefficient function estimation,
since 
$\| \hat{\beta}_{ij}(t) - \hat{\beta}_{i'j}(t) \|_\infty \le 2\underset{k}{\max}
\| \hat{\beta}_{kj}(t) - \beta_{kj}(t) \|_\infty$ converges to zero when $\mathcal{S}_j(i) = \mathcal{S}_j(i')$, and
$\| \hat{\beta}_{ij}(t) - \hat{\beta}_{i'j}(t) \|_\infty \geq 
\underset{\mathcal{S}_j(i) \neq \mathcal{S}_j(i')}{\min}
\| \beta_{ij}(t) - \beta_{i'j}(t) \|_\infty - 
2\underset{k}{\max}
\| \hat{\beta}_{kj}(t) - \beta_{kj}(t) \|_\infty>0$ for large sample when
$\mathcal{S}_j(i) \neq \mathcal{S}_j(i')$,
where {
$\mathcal{S}_j(i)$}
denotes the true subgroup index of the $i$-th subject with respect to the $j$-th covariate.
Consistency of HFLM is illustrated in  Supplementary Materials}. 
Theoretical results for other generalized outcomes could be derived in a similar manner.

\subsection{Theory for Pre-clustering}

In this subsection, we 
provide a theoretical guarantee of our implementation of  
the proposed pre-clustering 
and establish properties of the pre-clustering group identification for Gaussian reponses.
The detailed steps of implementation of the proposed pre-clustering  are provided in Supplementary Materials.

In Proposition \ref{thm: alg converge (SSE)} below, 
we provide the convergence of 
the pre-clustering algorithm which can be found in Supplementary Materials, 
requiring the following additional regularity condition (C7) which assumes that the number of true subgroups is much smaller than that of pre-clustering groups. 

{

(C7) The number of pre-clustering groups $K = O(1)$, and it satisfies that $\prod_{j=1}^p \mathcal{K}_j<<K$ for $j\in\{1,\dots,p\}$.
}

\begin{proposition}
\label{thm: alg converge (SSE)}
When Conditions (C1), (C2), (C5), (C6), and (C7) are satisfied, we have
$\operatorname{SSE}^{(l+1)}
< \operatorname{SSE}^{(l)}
$ 
for $ l=0,1,2, \dots$, 
where $l$ denotes the number of iterations in the pre-clustering algorithm, $\operatorname{SSE}^{(l)} = \sum_{i=1}^{n} (y_i - \hat{y}_i^{(l)})^2$, 
$\hat{y}_i^{(l)} = \tilde{\alpha}^{(l)} + \sum_{j=1}^p \int_0^T \tilde{\beta}_{(k)j}^{(l)}(t)X_{ij}(t) dt $, and $k\in \{1,\dots,K\}$ is 
the pre-clustering group index of the $i$-th subject at the $l$-th iteration. 
\end{proposition}
The strict decrease in SSE in Proposition \ref{thm: alg converge (SSE)} implies the convergence of the pre-clustering algorithm.

To demonstrate the consistency of the pre-clustering group identification, we introduce more notations and regularity conditions.
We denote the population distribution measure of responses as $P$, and denote the empirical measure as $P_n$.
For any probability measure $Q$ on $\mathbb{R}$ 
and any subset $\tilde{B}$ of $\mathbb{R}^{pL \times 1}$,
we define  $\Phi(\tilde{B},Q, \alpha) = \int \min_{
\tilde{b}\in \tilde{B} } \|Y- \alpha - \sum_{j=1}^p \gamma_j^T F_j^T \tilde{b}\|^2 Q(dY) $,
where $\gamma_j = \int_0^T X_j(t)\bm{B}(t)dt $ and
$F_j$ is defined similarly to $E_j$ in Equation (\ref{def_E_i}) with $n$ replaced by $p$.
When the responses follow a Gaussian distribution and ${\beta}_{(k),j}(t) \in S_{dM}$ for $k\in\{1,\dots,K\}, j\in\{1,\dots,p\}$, the empirical loss function in Equation (\ref{loss function: pre-clustering}) for the pre-clustering 
is equal to $\Phi(\tilde{B}(K), P_n, \alpha)$, where $\tilde{B}(K)$
consists of B-spline coefficients of $\beta_{(1)},\dots,\beta_{(K)}$.

We let $m_K(Q) = \inf\{\Phi(\tilde{B},Q, \alpha): \tilde{B} \text{ contains K or fewer points}, \alpha\in\mathbb{R}  \} $ denote the minimum of the loss function $\Phi$.
For a given $K$ and the sample size $n$, let 
$\tilde{B}_n = \tilde{B}_n(K) = \{\tilde{b}_{n(1)}, \dots, \tilde{b}_{n(K)}\}$ and $\tilde{\alpha}_n$ be the minimizer of $\Phi$ for $P_n$ such that $\Phi(\tilde{B}_n, P_n, \tilde{\alpha}_n) = m_K(P_n)$, 
where $\tilde{b}_{n(k)} = (\tilde{b}_{n,(k),1}^T, \dots, \tilde{b}_{n,(k),p}^T)^T$ for $k \in \{1, \dots, K\}$. 
Note that $\tilde{\alpha}_n $ and $\tilde{\beta}_{n(k)} = (\tilde{b}_{n,(k),1}^T \bm{B}(t),$ $\dots, \tilde{b}_{n,(k),p}^T \bm{B}(t))^T$ equal $ \tilde{\alpha}$ and $\tilde{\beta}_{(k)}$, respectively,  in Equation (\ref{minimizer: pre-clustering}) 
when we have Gaussian outcomes.
We also let 
$\Bar{B} = \Bar{B}(K) = \{ \Bar{b}_{(1)}, \dots, \Bar{b}_{(K)}\}$ and $\bar{\alpha}$ be the oracle versions of $\tilde{B}_n$ and $\tilde{\alpha}_n$, respectively, such that $\Phi(\Bar{B}, P, \Bar{\alpha}) = m_K(P)$, where $\Bar{b}_{(k)} = \{\Bar{b}_{(k),1}, \dots, \Bar{b}_{(k),p}\}$ for $k \in\{ 1, \dots, K\}$.

{
Moreover, we define an additional metric to evaluate how well the pre-clustering method captures a finer partition of the true subgroup structure. 
For the $j$-th covariate, we consider the following {Purity} measure:
\begin{equation}
\small 
\setlength\abovedisplayskip{3pt}
\setlength\belowdisplayskip{4pt}
\tilde{\mathcal{P}}_j(\tilde{\mathcal{G}})
=
\frac{1}{K}\sum_{k=1}^{K}
\frac{1}{\lvert \tilde{\mathcal{G}}_k\rvert}\;\max_{l \in \{1, \dots, \mathcal{K}_j \} }\Bigl\lvert\{\,i\in \tilde{\mathcal{G}}_k : \mathcal{S}_j(i)=l\}\Bigr\rvert, 
\quad j\in\{1,\dots,p\},
\label{eq: Purity for pre-clustering}
\end{equation}
where $| \cdot |$ measures the number of elements in the set.

This metric evaluates how well each pre-clustering group  $\tilde{\mathcal{G}}_k$ refines the true subgroup structure.
The value of $\tilde{\mathcal{P}}_j(\tilde{\mathcal{G}})$ lies in $[0,1]$. 
A value of  $1$ indicates that the pre-clustering groups form a perfect finer partition of the true subgroup structure, while smaller values indicate less accurate finer partitions.
}

The following are two additional regularity conditions.

(C8) For any $j\in \{1,\dots,K\},$ there exists unique $ \Bar{B}(j)$ and $\Bar{\alpha}$ \text{such that} $\Phi(\Bar{B}(j), P, \Bar{\alpha}) = m_j(P) $.


(C9) For any $j\in \{1,\dots,p\}$ and $l \in\{1,\dots,\mathcal{K}_j\} $, there exists a unique sub-sequence 
$\mathcal{A}_l \subset \{1,\dots, K \}$ and a constant $\tau>0$
such that for any
$k \in \mathcal{A}_l$ and $i\in \mathcal{G}_{j,l}  $,
we have $\| \left(\Bar{\beta}_{(k),j}(t) - {\beta}_{i,j}(t)\right)^2\|_\infty = O(1/n^\tau)$,  where  $\Bar{\beta}_{(k),j}(t) = \Bar{b}_{(k),j}^T \mathbf{B}(t)$.

Conditions (C8)  is analogous to assumptions in \citep{pollard1981strong}. 
Condition (C9) 
requires 
the oracle minimizer $\bar{\beta}_{(k),j} (t)$'s in the pre-clustering procedure to be close to  true coefficient functions.
Specifically,
for any true coefficient function $\beta_{ij} (t)$, there are some $\bar{\beta}_{(k),j} (t)$'s close to $\beta_{ij} (t)$ in terms of the infinity norm.

{
\begin{theorem}
\label{thm: pre-clustering}
When Conditions  (C2), (C5), (C6), (C7), (C8), and (C9) are satisfied and responses are Gaussian distributed,
we have
$ \Phi(\tilde{B}_n, P_n , \tilde{\alpha}_n) \xrightarrow{a.s.} m_K(P)$, and with probability one, the corresponding pre-clustering group identification $\tilde{\mathcal{G} }= \{ \tilde{\mathcal{G}}_{1},\dots, \tilde{\mathcal{G}}_{K} \}$ satisfies that 
$ \operatorname{Pr}\left(\tilde{\mathcal{P}}_j(\tilde{\mathcal{G}})=1\right) \rightarrow 1$ for $j \in \{1,\dots,p\}$.
\end{theorem}




The almost sure convergence of 
$\Phi(\tilde{B}_n, P_n, \tilde{\alpha}_n)$ in Theorem \ref{thm: pre-clustering} implies that $\tilde{B}_n \xrightarrow{a.s.} \Bar{B}$ under Condition (C8). 
This  indicates that, with probability one, the empirical estimator $\tilde{\beta}_{(1)}, \dots, \tilde{\beta}_{(K)}$ defined in Equation \eqref{minimizer: pre-clustering} converge to the oracle minimizer $\Bar{\beta}_{(1)}, \dots, \Bar{\beta}_{(K)}$ 
 since $\Phi(\tilde{B}, P_n, \alpha)$ is equivalent to the pre-clustering loss function for Gaussian responses, where $\Bar{\beta}_{(k)}=(\Bar{\beta}_{(k),1}(t),\dots,\Bar{\beta}_{(k),p}(t))^T$.
Theorem \ref{thm: pre-clustering} also demonstrates that, with probability one, the $i$-th subject and $i'$-th subject will not be estimated in the same pre-clustering group if $\mathcal{S}_j(i) \neq \mathcal{S}_j(i')$  for $i,i'\in \{1,\dots,n\}$, $j\in \{1,\dots,p \}$.
In other words, the proposed pre-clustering groups provide a finer partition of the true subgroup identification, and it will not break the true subgroup structure.
}

}

{\color{black}
\section{Simulation}
\label{section: simulation}

{
We conduct a comprehensive simulation study to evaluate the performance of our proposed GHFM and pre-clustering strategy under a variety of scenarios, in comparison with existing methods.
The existing approaches include the Homogeneous Smooth Functional Linear Model (HSFLM) \citep{cardot2003spline} that applies a single smooth functional linear regression to the entire dataset,
an oracle benchmark that applies HSFLM to each true subgroup, Functional Mixture Regression (FunMR) \citep{yao2011functional} that can be achieved 
via the \texttt{flexmix} package, and a Deviance-based K-means method (DevKM) that clusters the data with respect to deviance and then fits HSFLM within each cluster; for Gaussian responses, DevKM reduces to the Resi method  in \cite{sun2022subgroup}.

We evaluate the performance of each method using four criteria. Purity that is defined in Equation (\ref{eq: Purity for pre-clustering}) measures the proportion of pre-clusters containing subjects from exactly one true subgroup. It  evaluates the quality of the pre-clustering step. 
Normalized Mutual Information (NMI) quantifies the agreement between the estimated and true subgroup labels, and is defined as
{
\small
\[
\setlength\abovedisplayskip{3pt}
\setlength\belowdisplayskip{4pt}
\mathrm{NMI}(\mathcal{G}_{j}, \hat{\mathcal{G}}_j) = \frac{2 \cdot I(\mathcal{G}_{j}; \hat{\mathcal{G}}_j) } {H(\mathcal{G}_{j}) + H(\hat{\mathcal{G}}_j)},
\]
}where $\mathcal{G}_{j}$ denotes the true subgroup identification for the $j$-th covariate and $\hat{\mathcal{G}}_j$ denotes the estimated subgroup identification for the $j$-th covariate. 
Here, $I(\mathcal{G}_{j}; \hat{\mathcal{G}}_j)$ measures the mutual dependence between the true and estimated subgroup assignments, and $H(\cdot)$ measures the uncertainty (entropy) of a labeling. 
Therefore, NMI takes values in $[0,1]$, with $1$ indicating perfect agreement and $0$ indicating no mutual information between the two partitions.
The estimation accuracy of coefficient functions is assessed by the integrated squared error (ISE), which is defined as
\[
\small 
\setlength\abovedisplayskip{3pt}
\setlength\belowdisplayskip{4pt}
\text{ISE}(\hat{\beta}) = \frac{\| \hat{\boldsymbol{\beta}}(t) - \boldsymbol{\beta}(t) \|_2}{\| \boldsymbol{\beta}(t) \|_2}, \quad
\text{ where }
\| \boldsymbol{\beta}(t) \|_2 = \left( \sum_{k=1}^{K} \int_0^{T} \beta_k^2(t) dt \right)^{1/2}.
\]
Finally, we report the $p$-value associated with the hypotheses in (\ref{eq: H0 and Ha of testing effects in different groups}), which assess whether subgroup-specific functional effects differ significantly.

We focus on the single–covariate case; results for multiple covariates are similar and can be found in  Supplementary Materials. 
For each replication, we generate $n$ independent functional predictors $\{X_{i1}(t)\}_{i=1}^n$ on the domain $[0,1440]$, with $X_{i1}(t_k)\sim N(3,1)$ at $k=1,\dots,1440$ equally spaced time points. 
{
From the discrete observations \(\{X_{i1}(t_k)\}_{k=1}^{1440}\), each predictor is reconstructed as a smooth function $X_{i1}(t)$ via an approximation of a shared \(15\)-dimensional B\textendash spline basis $\boldsymbol{B}(t)$.}
We consider $\mathcal{K}_1\in\{2,4\}$ true subgroups and assign subjects to subgroups uniformly at random. 
We generate Gaussian type outcomes by 
\[
\setlength\abovedisplayskip{3pt}
\setlength\belowdisplayskip{4pt}
\small
y_i=\int_0^{1440} X_{i1}(t)\,\beta_{i1}(t)\,dt+\varepsilon_i, \quad \varepsilon_i\sim N(0,\sigma_\epsilon^2), \quad  \mathcal{S}_1(i)=k\in\{1,\dots,\mathcal{K}_1\}
\]
and  Bernoulli outcomes by $y_i\sim \text{Bernoulli}(p_i)$ where $p_i$ satisfies
\[
\setlength\abovedisplayskip{3pt}
\setlength\belowdisplayskip{4pt}
\small
\log\frac{p_i}{1-p_i}=\int_0^{1440} X_{i1}(t)\,\beta_{i1}(t)\,dt, \quad  \mathcal{S}_1(i)=k\in\{1,\dots,\mathcal{K}_1\}.
\]
The subgroup-specific functional effect $\beta_{i1}(t)$ is constructed under the following settings. 

\paragraph[Setting 1]{Setting 1: Random B-spline structure.} \label{case:bspline}
For each subject $i$ with $\mathcal{S}_1(i)=k\in\{1,\dots,\mathcal{K}_1\}$, we 
define the effect function by $\beta_{i1}(t)=\boldsymbol{b}_{i1}^\top \boldsymbol{B}(t)
$
with $\boldsymbol{b}_{i1}\sim N_L(\boldsymbol{\mu}_k,\sigma^2 I_L)$ and $L=15$.
The noise level $\sigma$ controls the separation between subgroups. For $\mathcal{K}_1=2$ we set $\boldsymbol{\mu}_1=(5,\dots,5)^\top$ and $\boldsymbol{\mu}_2=(-5,\dots,-5)^\top$; for $\mathcal{K}_1=4$ we additionally set $\boldsymbol{\mu}_3=(2,\dots,2)^\top$ and $\boldsymbol{\mu}_4=(-2,\dots,-2)^\top$.

\paragraph[Setting 2]{Setting 2: Structured polynomial forms.} \label{case:poly}
For each subject $i$ with $\mathcal{S}_1(i)=k\in\{1,\dots,\mathcal{K}_1\}$,
we define $\beta_{i1}(t)$ as follows: odd–numbered subgroups follow a linear trend $\beta_{i1}(t)=a_k+b_k t$ with $a_k=-3+0.5\cdot\frac{k-1}{2}$ and $b_k=0.005-0.002\cdot\frac{k-1}{2},$
while even–numbered subgroups follow a quadratic trend $\beta_{i1}(t)=A_k+B_k t+C_k t^2$ with $A_k=-1-0.5\Big(\frac{k}{2}-1\Big),  B_k=-0.005+0.001\Big(\frac{k}{2}-1\Big)$
and
$C_k=0.000005-0.000002\Big(\frac{k}{2}-1\Big).$


{
The differences between Setting~1 and Setting~2 are as follows.
In Setting~1, subgroup differences are primarily reflected in shifts of the mean level. 
In Setting~2, subgroup differences arise from distinct functional forms, such as linear versus quadratic shapes.
}




{
Tables~\ref{tab:simu-res-case1} and \ref{tab:simu-res-case2} show the simulation results for $\mathcal{K}_1 = 2$ with $L = 20$ B-spline basis functions under the proposed GHFM method. The results for $\mathcal{K}_1 = 4$ are reported in Supplementary Materials and are similar to the case with $\mathcal{K}_1 = 2$. Supplementary Materials also report results for varying $L$. We find that the proposed GHFM method remains stable once $L$ is sufficiently large. Increasing the number of basis functions can improve accuracy but leads to higher computational cost. A moderate choice such as $L = 15$ or $L = 20$ provides a good trade-off between performance and efficiency for the proposed method.

Tables~\ref{tab:simu-res-case1} and \ref{tab:simu-res-case2} show that our proposed GHFM achieves accurate subgroup recovery and coefficient estimation. 
With a large sample size ($n=10000$), the proposed GHFM performs very close to the oracle in both settings. 
In Setting~1 (Table~\ref{tab:simu-res-case1}), the GHFM attains NMI of $0.93$–$0.96$ and ISE of $0.32$–$0.52$, compared with the oracle’s perfect NMI of $1.00$ and lower ISE of $0.20$–$0.22$. 
The FunMR and DevKM (NMI $0.79$–$0.89$, ISE $0.63$–$0.71$) perform worse than the proposed GHFM, while HSFLM, which ignores heterogeneity of effect, suffers from very large ISE around $2.5$. 
A similar pattern appears in Setting~2 (Table~\ref{tab:simu-res-case2}), where GHFM achieves NMI above $0.92$ and ISE $0.36$–$0.57$, again close to oracle (NMI $1.00$, ISE $0.17$–$0.19$), outperforming FunMR and DevKM (NMI $0.77$–$0.88$, ISE $0.66$–$0.73$) and substantially better than HSFLM. 

\begin{table}
\small
\centering
\begin{adjustbox}{max width=1.0\textwidth}
\begin{tabular}{ll|llll|llll}
\toprule
       &            & \multicolumn{4}{c|}{\textbf{Gaussian Response}} & \multicolumn{4}{c}{\textbf{Bernoulli Response}} \\
$n$    & Method     & NMI & Purity & ISE  & $p$-value & NMI & Purity & ISE  & $p$-value \\
\midrule
\multirow{8}{*}{100}
& GHFM$^*$   & 0.92 & - & 0.58 & $<$0.05 & 0.90 & - & 0.88 & $<$0.05 \\
& GHFM\_10   & 0.91 & 0.93 & 0.62 & $<$0.05 & 0.89 & 0.92 & 0.92 & $<$0.05 \\
& GHFM\_20  & 0.90 & 0.93 & 0.67 & $<$0.05 & 0.89 & 0.91 & 0.95 & $<$0.05 \\
& Oracle    & 1.00 &  –   & 0.51 &  –   & 1.00 &  –   & 0.57 &  –   \\
& FunMR     & 0.77 &  –   & 1.02 &  –   & 0.72 &  –   & 1.16 &  –   \\
& DevKM     & 0.84 &  –   & 0.98 &  –   & 0.80 &  –   & 1.09 &  –   \\
& HSFLM     &  –   &  –   & 2.80 &  –   &  –   &  –   & 2.85 &  –   \\
\midrule
\multirow{8}{*}{10000}
& GHFM\_50   & 0.96 & 0.97 & 0.32 & $<$0.05 & 0.93 & 0.96 & 0.48 & $<$0.05 \\
& GHFM\_100  & 0.96 & 0.95 & 0.34 & $<$0.05 & 0.94 & 0.94 & 0.51 & $<$0.05 \\
& GHFM\_150  & 0.94 & 0.96 & 0.35 & $<$0.05 & 0.93 & 0.94 & 0.52 & $<$0.05 \\
& Oracle     & 1.00 &  –   & 0.20 &  –   & 1.00 &  –   & 0.22 &  –   \\
& FunMR      & 0.82 &  –   & 0.66 &  –   & 0.79 &  –   & 0.71 &  –   \\
& DevKM      & 0.89 &  –   & 0.63 &  –   & 0.87 &  –   & 0.69 &  –   \\
& HSFLM      &  –   &  –   & 2.65 &  –   &  –   &  –   & 2.54 &  –   \\
\bottomrule
\end{tabular}
\end{adjustbox}
\caption{Simulation results under Setting 1. 
{GHFM\_$K$ denotes the GHFM with $K$ pre-clustering groups and GHFM$^*$ indicates that GHFM is directly applied to all subjects.}
}
\label{tab:simu-res-case1}
\end{table}

\begin{table}
\small
\centering
\begin{adjustbox}{max width=1.0\textwidth}
\begin{tabular}{ll|llll|llll}
\toprule
       &            & \multicolumn{4}{c|}{\textbf{Gaussian Response}} & \multicolumn{4}{c}{\textbf{Bernoulli Response}} \\
$n$    & Method     & NMI & Purity & ISE  & $p$-value & NMI & Purity & ISE  & $p$-value \\
\midrule
\multirow{8}{*}{100}
& GHFM$^*$   & 0.90 & - & 0.64 & $<$0.05 & 0.88 & - & 0.92 & $<$0.05 \\
& GHFM\_10   & 0.88 & 0.91 & 0.68 & $<$0.05 & 0.88 & 0.91 & 0.96 & $<$0.05 \\
& GHFM\_20  & 0.88 & 0.91 & 0.70 & $<$0.05 & 0.87 & 0.90 & 0.96 & $<$0.05 \\
& oracle    & 1.00 &  –   & 0.53 &  –   & 1.00 &  –   & 0.58 &  –   \\
& FunMR     & 0.74 &  –   & 1.14 &  –   & 0.70 &  –   & 1.18 &  –   \\
& DevKM     & 0.82 &  –   & 1.07 &  –   & 0.78 &  –   & 1.11 &  –   \\
& HSFLM     &  –   &  –   & 2.65 &  –   &  –   &  –   & 2.77 &  –   \\
\midrule
\multirow{8}{*}{10000}
& GHFM\_50   & 0.94 & 0.96 & 0.36 & $<$0.05 & 0.93 & 0.95 & 0.57 & $<$0.05 \\
& GHFM\_100  & 0.92 & 0.94 & 0.38 & $<$0.05 & 0.92 & 0.95 & 0.57 & $<$0.05 \\
& GHFM\_150  & 0.93 & 0.94 & 0.38 & $<$0.05 & 0.92 & 0.93 & 0.57 & $<$0.05 \\
& oracle     & 1.00 &  –   & 0.17 &  –   & 1.00 &  –   & 0.19 &  –   \\
& FunMR      & 0.79 &  –   & 0.69 &  –   & 0.77 &  –   & 0.73 &  –   \\
& DevKM      & 0.88 &  –   & 0.66 &  –   & 0.86 &  –   & 0.70 &  –   \\
& HSFLM      &  –   &  –   & 2.37 &  –   &  –   &  –   & 2.46 &  –   \\
\bottomrule
\end{tabular}
\end{adjustbox}
\caption{Simulation results under Setting 2. {GHFM\_$K$ denotes the GHFM with $K$ pre-clustering groups and GHFM$^*$ indicates that GHFM is directly applied to all subjects.}
}
\label{tab:simu-res-case2}
\end{table}

When the sample size is smaller ($n=100$), the performance of GHFM slightly declines but is still better than other existing methods except for oracle. 
In Setting~1, GHFM achieves NMI $0.89$–$0.92$ and ISE $0.58$–$0.95$, close to oracle (NMI $1.00$, ISE $0.51$-$0.57$) and better than FunMR and DevKM (NMI $0.72$–$0.84$, ISE $0.98$-$1.16$). 
HSFLM again shows severe misspecification with ISE around $2.8$. 
In Setting~2, GHFM maintains NMI $0.87$–$0.90$ and ISE $0.64$–$0.96$, reasonably close to oracle (NMI $1.00$, ISE $0.53$–$0.58$). 
FunMR and DevKM remain worse (NMI $0.70$–$0.82$, ISE $1.07$–$1.18$).
Overall, the results across both settings for both types of response confirm that GHFM improves with larger sample sizes and  outperforms these existing methods.

In Tables~\ref{tab:simu-res-case1}--\ref{tab:simu-res-case2}, GHFM\_K denotes the GHFM with \(K\) pre-clustering groups, and GHFM\(^*\) denotes the GHFM applied directly to all subjects.
For $n=100$, we use $K\in\{10,20\}$, and for $n=10000$ we take $K\in\{50,100,150\}$.  
When \(n=100\), GHFM\(^*\), GHFM\_10, and GHFM\_{20} perform similarly; for example, their NMI values are all around \(0.9\) in Table \ref{tab:simu-res-case1}.
Moreover, Purity remains above \(0.9\) in all settings, indicating that the pre-clustering step provides high-quality, finer partitions of the true subgroups.
In addition, we observe that varying $K$ does not have much influence as long as it substantially exceeds the true subgroup number. 
For example, in Setting~1 with $n=10000$, GHFM\_50, GHFM\_100, and GHFM\_150 produce similar results (NMI $0.93$–$0.96$, ISE $0.32$–$0.52$), and the same holds in Setting~2 (NMI $0.92$–$0.94$, ISE $0.36$–$0.57$). 
Finally, all reported $p$-values are below $0.05$, which provides strong statistical evidence for the presence of subgroup-specific functional differences.


}

}
}

{
\section{Real Data Application}
\label{section: dementia application}
\subsection{The UKB Data}
In this section, we apply the proposed and existing methods to physical activity and health outcomes in the UKB study, a population-based cohort of over 500{,}000 participants \citep{sudlow2015uk}. 
We focus on accelerometry data which are widely used as the gold standard for objectively measuring physical activity \citep{strath2013guide}. Between February 2013 and December 2015, 103,720 participants wore an Axivity AX3 wrist-mounted triaxial accelerometer for one week. This device recorded continuous triaxial acceleration at 100 Hz with a dynamic range of $\pm$8 mg \citep{doherty2017large}, as documented in UKB field 90001. Raw signals were aggregated into five-second epochs and the Euclidean norm of the triaxial acceleration was computed \citep{sabia2014association} after the gravitational and noise components were removed \citep{da2014physical, van2011estimation}. 

Furthermore, we calculate average activity levels on a minute basis using the five-second epoch acceleration data \citep{karas2022comparison}. 
This calculation produces a vector with 1440 elements for each participant on each day, where each element corresponds to the average activity level for that specific minute, ranging from 00:00 (first element) to 23:59 (last element). 
We mainly focus on
the middle six days in the week of data collection
since the 24-hour cycles in the first and last days of the week are incomplete.
We remove 7,277 subjects deemed of suboptimal quality by the UKB team
\citep{leroux2021quantifying}, and the final sample size of physical activity data is  96,443.

Figure~\ref{fig: raw PA vs Smooth func} shows four subjects with distinct daily activity patterns. 
For example, Subject~1 shows higher activity  in the afternoon and evening, while Subject~3 remains low for most of the day and only increases during the lunch time.
In addition, Figure~\ref{fig: raw PA vs Smooth func} compares the raw minute-level data with the smoothed curves. Using 50 order-4 B-spline basis functions yields a good approximation that preserves minute-level fluctuations, so we adopt this setting in the real-data application.

\begin{figure}[h]
    \centering
    \includegraphics[width=0.65\textwidth]{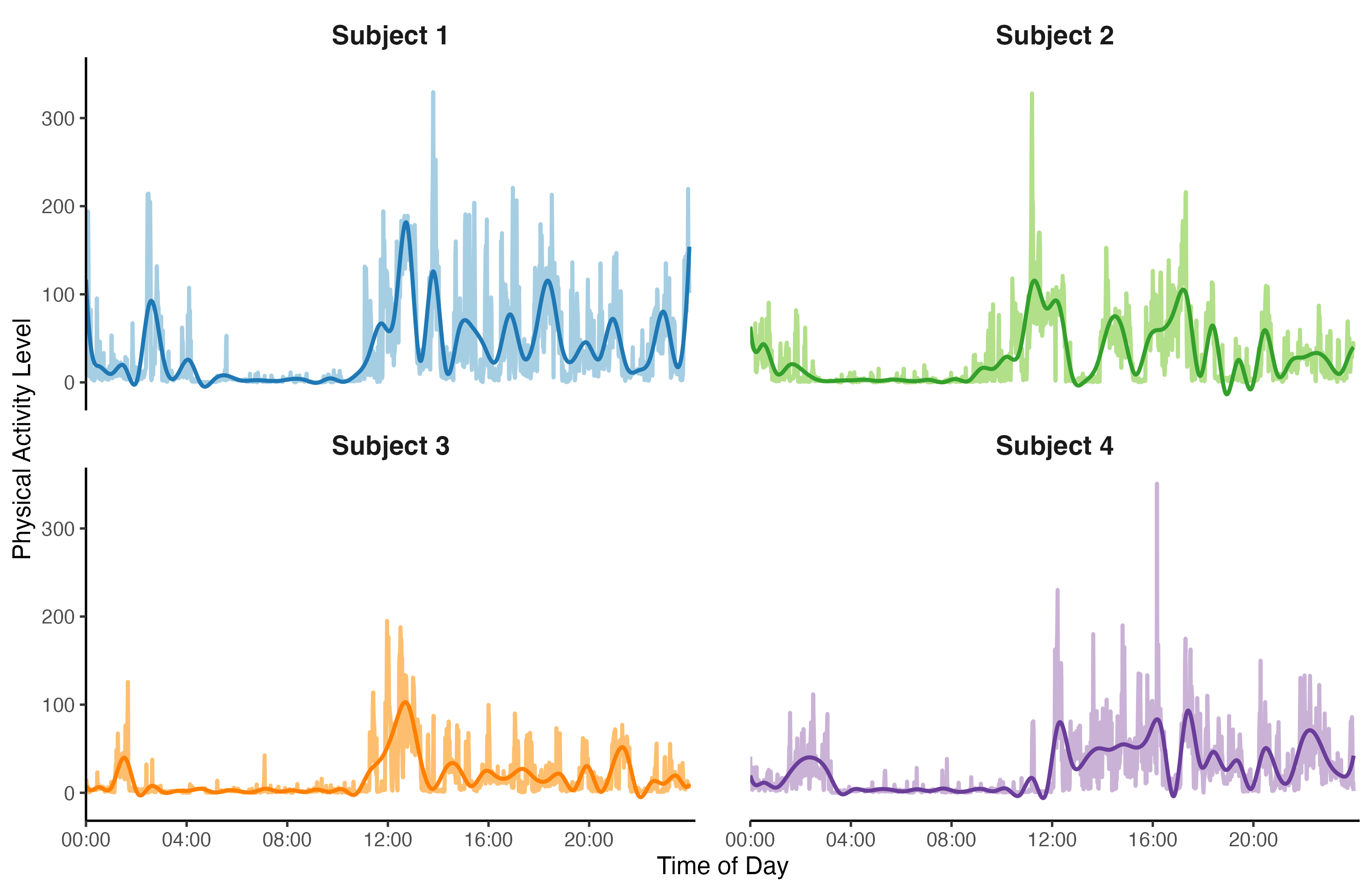}
    \caption{Comparison of raw and smoothed physical activity signals for four subjects, based on minute-level data from 0:00 to 23:59. The smoothed curves are obtained using 50 B-spline basis functions. Each panel represents one subject, with distinct color schemes for raw and smoothed curves.}
    \label{fig: raw PA vs Smooth func}  
\end{figure}


Dementia is a clinical syndrome of progressive cognitive decline beyond what is expected from normal ageing; it includes Alzheimer’s disease, vascular dementia, frontotemporal dementia, and other subtypes  \citep{duong2017dementia}. 
In the UKB study, we define an all-cause dementia indicator (0/1) as follows: a participant is coded 1 if they have ever been diagnosed with dementia in any of three sources — baseline self-report, hospital admission records, or death registry — otherwise coded 0 \citep{wilkinson2019identifying}.

\subsection{Application Results}

In this subsection, we apply the proposed GHFM to the  UKB physical activity and dementia data.
Since the sample size (96,443) is large, we first apply a pre-clustering method with $K=100$, and then apply the proposed HFLM to the data {where we use $L=15$ B-spline basis functions.} 
For comparison, in addition to the methods introduced in Section \ref{section: simulation}, we also include homogeneous logistic regression (LogiReg) to evaluate whether using a functional form can capture more information and yield better results.
Unlike the simulation study, where known true subgroup labels allow metrics such as NMI or ISE to be computed, in the real data setting the true subgroup structure is unknown and these measures cannot be applied.
{
Instead, we use the physical activity data from the first entire day as the training predictor $X_{1}(t)$ and the dementia diagnosis as the outcome $Y$ to fit our proposed method, and we reserve the subsequent five entire days as the testing data.
 }

We evaluate performance of each method using predictive false negative rate (PFNR), predictive false positive rate (PFPR), and predictive area under the curve (PAUC). 
Specifically, PFNR and PFPR are defined as the average false negative rate (FNR) and false positive rate (FPR) across the testing days, respectively, while PAUC represents the average area under the ROC curves across the five testing days.
{
In addition, we report the $p$-value $p_{\text{sub}}$ for testing \eqref{eq: H0 and Ha of testing effects in different groups} on the estimated subgroups.
A smaller value of \(p_{\text{sub}}\) provides stronger evidence for the existence of heterogeneity of effect.
}

The PFNRs, PFPRs, PAUCs, and \(p_{\text{sub}}\) for different methods are presented in Table \ref{table:pd_pred}.
The proposed method reduces the PFNR of each existing method by at least $40\%$ while maintaining a low PFPR.
Although PFPR of our method is slightly higher than that of other methods, it remains very close to 0 at a value of 0.0025.
The proposed method also achieves a higher PAUC than existing methods, with an improvement of at least 18\%.

In addition to the predictive metrics, the small value of \(p_{\text{sub}} = 0.0004\) indicates that the estimated subgroups have significantly different functional effects. 
This result highlights the importance of accounting for effect heterogeneity in analyzing large-scale mobile health data, as ignoring it may lead to the loss of valuable information.

Indeed, methods such as HSFLM and LogiReg yield a PFNR close to 1, likely due to  imbalance in the dementia outcome (only \(1.7\%\) of subjects are dementia cases) and their inability to capture heterogeneity of effects.
By contrast, our method, which explicitly considers heterogeneity, identifies a subgroup with a relatively higher case rate: \(2.82\%\) of individuals in Subgroup~2 are dementia cases (see Table~\ref{tab: env var comparison for dementia}), facilitating case detection and leading to a lower PFNR. {In addition, Table~\ref{tab: env var comparison for dementia} shows that, compared with other heterogeneous methods such as DevKM and FunMR, our proposed method also achieves better performance. This is likely because it attains more accurate subgroup identification with efficient computation and does not require predetermining the number of subgroups.
}


\begin{table}
\small
\centering
\begin{tabular}{lccccc}
\toprule
 & Proposed & HSFLM & LogiReg & FunMR & DevKM  \\
\midrule
PFNR & \textbf{0.2951} & 0.9236 & 1.0000 & 0.4169 & 0.4851  \\
PFPR & \textbf{0.0025} & 0.0007 & 0.0006 & 0.0062 & 0.0065  \\
PAUC & \textbf{0.8361} & 0.5914 & 0.5899 & 0.7120 & 0.7018  \\
$p_{\text{sub}}$ & 0.0004 & -- & -- & -- & --  \\
\bottomrule
\end{tabular}
\caption{Predictive performance on five future days for Dementia disease. 
PFNR/PFPR: predictive false negative/positive rates; PAUC: predictive area under ROC; \(p_{\text{sub}}\)    is the $p$-value testing subgroup-specific effects,
and \(p_{\text{ent}}\) is the $p$-value testing heterogeneity of effect for the entire dataset. }
\label{table:pd_pred}
\end{table}

Our proposed method identifies three distinct subgroups for dementia outcome.
Figure~\ref{fig:dementia_coef_panel} displays the estimated coefficient functions for all three subgroups, with the time axis spanning a full day from 00:00 to 24:00. 
{
The coefficient functions are non-positive for the majority of the daytime across all three subgroups. 
Suppose that we consider 06{:}00–19{:}30 as daytime, the estimated coefficient function is non-positive for 99.99\% of the daytime in Subgroup~1, 82.61\% in Subgroup~2, and 63.63\% in Subgroup~3. 
This aligns with existing evidence that physical activity tends to mitigate the risk of dementia \citep{iso2022physical, del2022association}.
}

Subgroup 2 is the subgroup with 2.82\% of the subjects with dementia, which is more than that of other subgroups. As shown in  Figure~\ref{fig:dementia_coef_panel}, the coefficient function of this subgroup exhibits a distinct pattern compared to other subgroups. Its coefficient function is positive between 11:00 PM and 4:00 AM with a relatively large magnitude, which indicates a positive relationship between nighttime activity and dementia risk. This positive relationship is possibly due to that individuals diagnosed with or at high risk for dementia often have sleep problem, leading to frequent nighttime awakenings and, consequently, increased physical activity during the night \citep{lysen2020actigraphy, lim2013sleep}.
We also observe that the coefficient function for Subgroup 2 is more negative from 8:00 AM to 12:00 PM than the remaining time within a day, suggesting that physical activity in the morning may reduce the risk of dementia for individuals in this subgroup more than other time \citep{ning2025accelerometer}.
Therefore, for people in  this subgroup, reducing late-night activity and instead doing suitable activity after waking may be advisable.

Subgroup~1 includes over 80\% of participants, so it likely reflects the typical pattern for most people in this dataset.
As shown in Figure~\ref{fig:dementia_coef_panel}, its coefficient function is non-positive for most of the day and becomes  negative from about 6:00~AM to 12:00~PM, reaching its minimum near 8:00~AM.
This may suggests that, for most normal people, physical activity does not increase dementia risk, and activity in the morning is associated with a lower risk of dementia \citep{zhang2023effect}.
Subgroup~3 shows an intermediate coefficient-function pattern between Subgroups~1 and~2.
Its coefficient function fluctuates around zero, with only brief positive intervals.
Even so, it tends to be negative in the morning, indicating that activity in the morning may also be associated with a lower dementia risk for this subgroup.

\begin{figure}[h]
\centering
\includegraphics[width=0.65\textwidth]{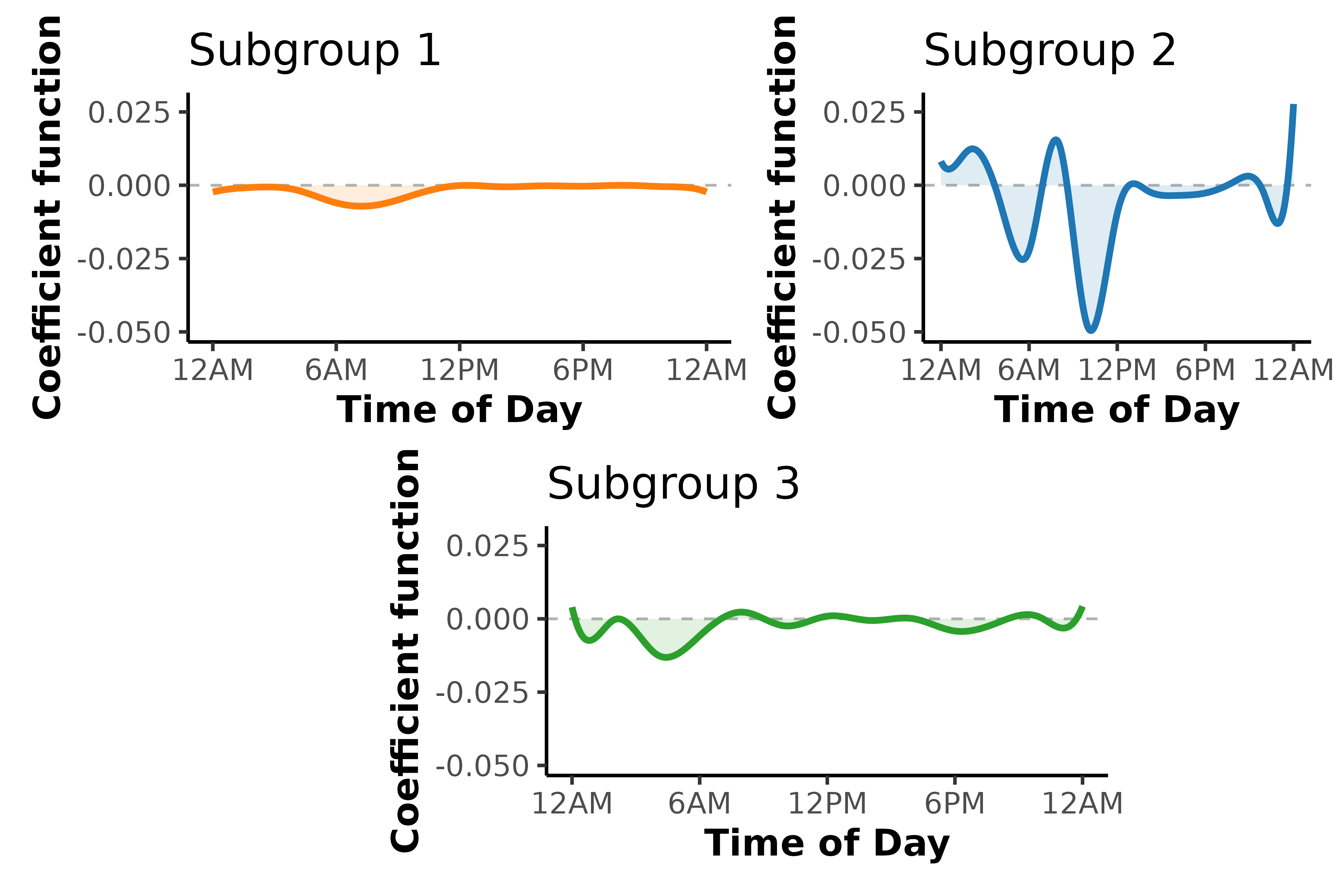}
\caption{Estimated coefficient functions for Dementia data.}
\label{fig:dementia_coef_panel}
\end{figure}

To acquire a deeper understanding of these subgroups, we conduct one-way ANOVAs for continuous variables and Chi–square tests for categorical variables to assess differences across the three subgroups in a broad set of UKB environmental variables. 
Table~\ref{tab: env var comparison for dementia}  summarizes some of the findings.


{
Table~\ref{tab: env var comparison for dementia} indicates that individuals in Subgroup~2 may be at higher risk of developing dementia.
There are a few factors that are related to increased risk of dementia: alcohol intake \citep{wiegmann2020alcohol, gulpers2016anxiety}, greater sedentary time \citep{raichlen2023sedentary}, 
mental unhealthy \citep{gulpers2016anxiety}, 
and insomnia \citep{meng2025insomnia}.
In fact, average weekly spirits intake is 2.45 in Subgroup~2 vs.~1.41/1.35 in Subgroups~1/3 ($p$ value $=0.031$), and the alcohol intake frequency is high in Subgroup~2. 
Subgroup~2 also reports more sedentary screen time (TV viewing 3.04 vs.~2.27/2.19 hours; $p$ value $=0.011$).
Moreover, Subgroup~2 reports higher anxiety and more frequent sleep problems; in particular, neuroticism is higher (3.97 vs.~3.90/3.81; $p$ value $=0.017$), and the prevalence of sleeplessness/insomnia is also higher (3.33\% vs.~2.47\%/2.28\%; $p$ value $=0.0132$).
In addition, other differences, such as a higher proportion reporting recent illness/injury/bereavement/stress (1.88\% vs.~1.12\%/1.06\%; $p$ value $=0.0414$) and slightly more risk-taking (27.35\% vs.~24.30\%/25.11\%; $p$ value $=0.041$), may also increase the risk.

}

\begin{table}[H]
\small
\centering
\begin{tabular}{>{\raggedright\arraybackslash}p{6.2cm}cccc}
\toprule
 & Subgroup 1 & Subgroup 2 & Subgroup 3 & p-value \\
 & (n = 83,121) & (n = 1,382) & (n = 11,940) & \\
\midrule
\textbf{Dementia, $N$ (\%)} & 1,471 (1.77\%) & 39 (2.82\%) & 184 (1.54\%) & \\
\addlinespace
\multicolumn{5}{l}{\emph{Continuous variables (group mean)}}\\
Exposure to tobacco smoke outside home 
& 0.06 & 0.94 & 0.07 & $<0.001$ \\
Time spent watching television (TV) 
& 2.27 & 3.04 & 2.19 & $0.011$ \\
Neuroticism score 
& 3.90 & 3.97 & 3.81 & $0.017$ \\
Average weekly spirits intake 
& 1.41 & 2.45 & 1.35 & $0.031$ \\
\midrule
\multicolumn{5}{l}{\emph{Categorical variables (group proportion of the most group-different level)}}\\
Alcohol intake frequency 
& 1.66\% & 2.46\% & 1.52\% & 0.0028 \\
Guilty feelings 
& 27.42\% & 30.82\% & 26.47\% & 0.0086 \\
Fed-up feelings 
& 9.91\% & 11.29\% & 9.57\% & 0.0098 \\
Sleeplessness / insomnia 
& 2.47\% & 3.33\% & 2.28\% & 0.0132 \\
Ever unenthusiastic/disinterested for a whole week 
& 23.51\% & 21.78\% & 23.97\% & 0.018 \\
Ever manic/hyper for 2 days 
& 32.46\% & 33.57\% & 32.93\% & 0.0276 \\
Plays computer games 
& 27.11\% & 28.36\% & 27.08\% & 0.0286 \\
Cheese intake 
& 15.88\% & 16.86\% & 16.34\% & 0.0314 \\
Leisure/social activities 
& 2.03\% & 2.75\% & 1.83\% & 0.0402 \\
Risk taking 
& 24.30\% & 27.35\% & 25.11\% & 0.041 \\
Illness, injury, bereavement, stress in last 2 years 
& 1.12\% & 1.88\% & 1.06\% & 0.0414 \\
\bottomrule
\end{tabular}
\caption{Characteristics of subgroups and differences in environmental variables for dementia data. 
The first row reports dementia case counts and percentages. 
For continuous variables, cell values are group means with p-values from one-way ANOVA. 
For categorical variables, cell values are the proportions (\%) of the most group-different category with p-values from chi-square tests.}
\label{tab: env var comparison for dementia}
\end{table}

Subgroup~1  generally shows more favorable values on these variables (lower TV viewing, lower spirits intake, lower secondhand–smoke exposure, and slightly lower neuroticism), and Subgroup~3 tends to fall between the other two on many measures (such as spirits intake 1.35 at baseline visit and relatively low TV time), which matches the coefficient–function pattern described earlier.

}

\section{Discussion}
\label{section: discussion and conclusion}
{
Our main contribution is to enhance the understanding of the potential heterogeneous functional effect of physical activity on diseases.
Specifically, we propose a GHFM within the generalized
FDA framework to capture this heterogeneous functional effect, develop a pre-clustering method to improve computational efficiency for large-scale data, 
and introduce a hypothesis testing procedure to validate whether the identified subgroups or the entire data exhibit heterogeneous functional effects.
The proposed methods are novel in their ability to handle large mobile health datasets and to identify subgroups with time-varying heterogeneous effects without predefining the number of subgroups, ensuring accurate insights and reduced computational cost.
It supports both continuous and generalized outcomes, making it versatile for real-world applications, and demonstrates superior future-day prediction accuracy over existing methods in large datasets such as the UKB study.

In the future, we could extend this framework to an online setting so that it can account for streaming physical activity data.
Moreover, exploring the nature of the subgroups identified by our method in more depth is also valuable. We may investigate factors associated with these subgroups, including potential correlations between genetic variables and subgroup classifications.
Finally, our methods could also be applied to other large-scale datasets, such as the All of Us dataset \citep{all2019all}, for further discoveries.
}

\section*{Supplementary Materials}
 Algorithms, proofs, additional simulation results, and additional details on the real data application can be found in the Supplementary Materials. 

\section*{Acknowledgments}
We gratefully acknowledge support  from the UK Biobank study and the Purdue Rosen Center For Advanced Computing (RCAC). This work was also supported by the National Science Foundation under Grant DMS 2210860.
{
We acknowledge that ChatGPT-4 and ChatGPT-5 (OpenAI) were used for coding assistance and vocabulary checks in the preparation of this manuscript. The authors take full responsibility for the content and integrity of the final work.
}


\section*{Disclosure Statement}
The authors report that there are no competing interests to declare.

\section*{Data Availability Statement}
The data that support the findings of this study are at \url{https://www.ukbiobank.ac.uk/}.
UK Biobank has obtained research ethics approval from the North West Multi‐centre Research Ethics Committee. All participants provided a written informed consent at recruitment.

\begingroup
\setstretch{1.1}  
\small
\footnotesize
\bibliographystyle{Chicago}
\bibliography{hetero_ref_upper_force.bib}

@article{shen1998local,
  title={{Local Asymptotics for Regression Splines and Confidence Regions}},
  author={Shen, Xiaotong and Wolfe, Douglas A. and Zhou, Shanggang},
  journal={The Annals of Statistics},
  volume={26},
  number={5},
  pages={1760--1782},
  year={1998},
  publisher={Institute of Mathematical Statistics}
}

@book{bosq2000linear,
  title={{Linear Processes in Function Spaces: Theory and Applications}},
  author={Bosq, Denis},
  volume={149},
  year={2000},
  publisher={New York: Springer}
}

@article{pollard1981strong,
  title={{Strong Consistency of K-Means Clustering}},
  author={Pollard, David},
  journal={The Annals of Statistics},
  volume = {9(1)},
  pages={135--140},
  year={1981},
  publisher={JSTOR}
}

@article{zhou2013functional,
  title={{Functional Linear Model With Zero-Value Coefficient Function at Sub-Regions}},
  author={Zhou, Jianhui and Wang, Nae-Yuh and Wang, Naisyin},
  journal={Statistica Sinica},
  volume={23},
  number={1},
  pages={25},
  year={2013},
  publisher={NIH Public Access}
}

@article{shim2023wearable,
  title={{Wearable-Based Accelerometer Activity Profile as Digital Biomarker of Inflammation, Biological Age, and Mortality Using Hierarchical Clustering Analysis in NHANES 2011--2014}},
  author={Shim, Jinjoo and Fleisch, Elgar and Barata, Filipe},
  journal={Scientific Reports},
  volume={13},
  number={1},
  pages={9326},
  year={2023},
  publisher={Nature Publishing Group UK London}
}

@article{huang2022sleep,
  title={{Sleep and Physical Activity in Relation to All-Cause, Cardiovascular Disease, and Cancer Mortality Risk}},
  author={Huang, Bo-Huei and Duncan, Mitch J and Cistulli, Peter A and Nassar, Natasha and Hamer, Mark and Stamatakis, Emmanuel},
  journal={British Journal of Sports Medicine},
  volume={56},
  number={13},
  pages={718--724},
  year={2022},
  publisher={BMJ Publishing Group Ltd and British Association of Sport and Exercise Medicine}
}

@book{garzon2017benefits,
  title={{The Benefits of Strength Training and Tips for Getting Started}},
  author={Garzon, Raquel Cristina},
  year={2017},
  publisher={College of Agricultural, Consumer and Environmental Sciences}
}

@article{ghosal2022scalar,
  title={{Scalar on Time-By-Distribution Regression and Its Application for Modelling Associations Between Daily-Living Physical Activity and Cognitive Functions in Alzheimer’s Disease}},
  author={Ghosal, Rahul and Varma, Vijay R and Volfson, Dmitri and Urbanek, Jacek and Hausdorff, Jeffrey M and Watts, Amber and Zipunnikov, Vadim},
  journal={Scientific Reports},
  volume={12},
  number={1},
  pages={11558},
  year={2022},
  publisher={Nature Publishing Group UK London}
}

@article{ekelund2019dose,
  title={{Dose-Response Associations Between Accelerometry Measured Physical Activity and Sedentary Time and All-Cause Mortality: Systematic Review and Harmonised Meta-Analysis}},
  author={Ekelund, Ulf and Tarp, Jakob and Steene-Johannessen, Jostein and Hansen, Bj{\o}rge H and Jefferis, Barbara and Fagerland, Morten W and Whincup, Peter and Diaz, Keith M and Hooker, Steven P and Chernofsky, Ariel and others},
  journal={British Medical Journal},
  volume={366},
  pages={l4570},
  year={2019},
  publisher={British Medical Journal Publishing Group}
}

@article{barker2019physical,
  title={{Physical Activity of UK Adults With Chronic Disease: Cross-Sectional Analysis of Accelerometer-Measured Physical Activity in 96,706 UK Biobank Participants}},
  author={Barker, Joseph and Smith Byrne, Karl and Doherty, Aiden and Foster, Charlie and Rahimi, Kazem and Ramakrishnan, Rema and Woodward, Mark and Dwyer, Terence},
  journal={International Journal of Epidemiology},
  volume={48},
  number={4},
  pages={1167--1174},
  year={2019},
  publisher={Oxford University Press}
}

@article{pearce2022association,
  title={{Association Between Physical Activity and Risk of Depression: A Systematic Review and Meta-Analysis}},
  author={Pearce, Matthew and Garcia, Leandro and Abbas, Ali and Strain, Tessa and Schuch, Felipe Barreto and Golubic, Rajna and Kelly, Paul and Khan, Saad and Utukuri, Mrudula and Laird, Yvonne and others},
  journal={JAMA Psychiatry},
  volume={79},
  number={6},
  pages={550--559},
  year={2022},
  publisher={American Medical Association}
}

@article{mahindru2023role,
  title={{Role of Physical Activity on Mental Health and Well-Being: A Review}},
  author={Mahindru, Aditya and Patil, Pradeep and Agrawal, Varun},
  journal={Cureus},
  volume={15},
  number={1},
  year={2023},
  publisher={Cureus Inc.}
}

@article{all2019all,
  title={{The “All of Us” Research Program}},
  author={Denny, Joshua C. and Rutter, Jonathan L. and Goldstein, David B. and Philippakis, Anthony and Smoller, Jordan W. and Jenkins, Graham and others},
  journal={New England Journal of Medicine},
  volume={381},
  number={7},
  pages={668--676},
  year={2019},
  publisher={Mass Medical Soc}
}

@article{althoff2017large,
  title={{Large-Scale Physical Activity Data Reveal Worldwide Activity Inequality}},
  author={Althoff, Tim and Sosi{\v{c}}, Rok and Hicks, Jennifer L and King, Abby C and Delp, Scott L and Leskovec, Jure},
  journal={Nature},
  volume={547},
  number={7663},
  pages={336--339},
  year={2017},
  publisher={Nature Publishing Group UK London}
}

@article{roig2016time,
  title={{Time-Dependent Effects of Cardiovascular Exercise on Memory}},
  author={Roig, Marc and Thomas, Richard and Mang, Cameron S and Snow, Nicholas J and Ostadan, Fatemeh and Boyd, Lara A and Lundbye-Jensen, Jesper},
  journal={Exercise and Sport Sciences Reviews},
  volume={44},
  number={2},
  pages={81--88},
  year={2016},
  publisher={LWW}
}

@article{rosmalen2012revealing,
  title={{Revealing Causal Heterogeneity Using Time Series Analysis of Ambulatory Assessments: Application to the Association Between Depression and Physical Activity After Myocardial Infarction}},
  author={Rosmalen, Judith GM and Wenting, Angela MG and Roest, Annelieke M and de Jonge, Peter and Bos, Elisabeth H},
  journal={Psychosomatic Medicine},
  volume={74},
  number={4},
  pages={377--386},
  year={2012},
  publisher={LWW}
}

@article{sabia2014association,
  title={{Association Between Questionnaire-And Accelerometer-Assessed Physical Activity: The Role of Sociodemographic Factors}},
  author={Sabia, S{\'e}verine and van Hees, Vincent T and Shipley, Martin J and Trenell, Michael I and Hagger-Johnson, Gareth and Elbaz, Alexis and Kivimaki, Mika and Singh-Manoux, Archana},
  journal={American Journal of Epidemiology},
  volume={179},
  number={6},
  pages={781--790},
  year={2014},
  publisher={Oxford University Press}
}

@article{da2014physical,
  title={{Physical Activity Levels in Three Brazilian Birth Cohorts as Assessed With Raw Triaxial Wrist Accelerometry}},
  author={da Silva, Inacio CM and van Hees, Vincent T and Ramires, Virg{\'\i}lio V and Knuth, Alan G and Bielemann, Renata M and Ekelund, Ulf and Brage, Soren and Hallal, Pedro C},
  journal={International Journal of Epidemiology},
  volume={43},
  number={6},
  pages={1959--1968},
  year={2014},
  publisher={Oxford University Press}
}

@article{van2011estimation,
  title={{Estimation of Daily Energy Expenditure in Pregnant and Non-Pregnant Women Using a Wrist-Worn Tri-Axial Accelerometer}},
  author={van Hees, Vincent T and Renstr{\"o}m, Frida and Wright, Antony and Gradmark, Anna and Catt, Michael and Chen, Kong Y and L{\"o}f, Marie and Bluck, Les and Pomeroy, Jeremy and Wareham, Nicholas J and others},
  journal={PLOS ONE},
  volume={6},
  number={7},
  pages={e22922},
  year={2011},
  publisher={Public Library of Science San Francisco, USA}
}

@article{ma2017concave,
  title={{A Concave Pairwise Fusion Approach to Subgroup Analysis}},
  author={Ma, Shujie and Huang, Jian},
  journal={Journal of the American Statistical Association},
  volume={112},
  number={517},
  pages={410--423},
  year={2017},
  publisher={Taylor \& Francis}
}

@article{claeskens2009asymptotic,
  title={{Asymptotic Properties of Penalized Spline Estimators}},
  author={Claeskens, Gerda and Krivobokova, Tatyana and Opsomer, Jean D},
  journal={Biometrika},
  volume={96},
  number={3},
  pages={529--544},
  year={2009},
  publisher={Oxford University Press}
}

@article{xue2010consistent,
  title={{Consistent Model Selection for Marginal Generalized Additive Model for Correlated Data}},
  author={Xue, Lan and Qu, Annie and Zhou, Jianhui},
  journal={Journal of the American Statistical Association},
  volume={105},
  number={492},
  pages={1518--1530},
  year={2010},
  publisher={Taylor \& Francis}
}

@article{leroux2021quantifying,
  title={{Quantifying the Predictive Performance of Objectively Measured Physical Activity on Mortality in the UK Biobank}},
  author={Leroux, Andrew and Xu, Shiyao and Kundu, Prosenjit and Muschelli, John and Smirnova, Ekaterina and Chatterjee, Nilanjan and Crainiceanu, Ciprian},
  journal={The Journals of Gerontology: Series A},
  volume={76},
  number={8},
  pages={1486--1494},
  year={2021},
  publisher={Oxford University Press US}
}

@article{rowlands2021association,
  title={{Association Between Accelerometer-Assessed Physical Activity and Severity of COVID-19 in UK Biobank}},
  author={Rowlands, Alex V and Dempsey, Paddy C and Gillies, Clare and Kloecker, David E and Razieh, Cameron and Chudasama, Yogini and Islam, Nazrul and Zaccardi, Francesco and Lawson, Claire and Norris, Tom and others},
  journal={Mayo Clinic Proceedings: Innovations, Quality \& Outcomes},
  volume={5},
  number={6},
  pages={997--1007},
  year={2021},
  publisher={Elsevier}
}

@article{yao2011functional,
  title={{Functional Mixture Regression}},
  author={Yao, Fang and Fu, Yuejiao and Lee, Thomas CM},
  journal={Biostatistics},
  volume={12},
  number={2},
  pages={341--353},
  year={2011},
  publisher={Oxford University Press}
}

@article{zhang2022subgroup,
  title={{Subgroup Analysis for High-Dimensional Functional Regression}},
  author={Zhang, Xiaochen and Zhang, Qingzhao and Ma, Shuangge and Fang, Kuangnan},
  journal={Journal of Multivariate Analysis},
  volume={192},
  pages={105100},
  year={2022},
  publisher={Elsevier}
}

@article{sun2022subgroup,
  title={{Subgroup Analysis for the Functional Linear Model}},
  author={Sun, Yifan and Liu, Ziyi and Wang, Wu},
  journal={arXiv Preprint arXiv:2211.15051},
  year={2022}
}

@article{doherty2017large,
  title={{Large-Scale Population Assessment of Physical Activity Using Wrist-Worn Accelerometers: The UK Biobank Study}},
  author={Doherty, Aiden and Jackson, Dan and Hammerla, Nils and Pl{\"o}tz, Thomas and Olivier, Patrick and Granat, Malcolm H and White, Tom and Van Hees, Vincent T and Trenell, Michael I and Owen, Christoper G and others},
  journal={PLOS ONE},
  volume={12},
  number={2},
  pages={e0169649},
  year={2017},
  publisher={Public Library of Science}
}

@article{sudlow2015uk,
  title={{UK Biobank: An Open Access Resource for Identifying the Causes of a Wide Range of Complex Diseases of Middle and Old Age}},
  author={Sudlow, Cathie and Gallacher, John and Allen, Naomi and Beral, Valerie and Burton, Paul and Danesh, John and Downey, Paul and Elliott, Paul and Green, Jane and Landray, Martin and others},
  journal={PLoS Medicine},
  volume={12},
  number={3},
  pages={e1001779},
  year={2015},
  publisher={Public Library of Science}
}

@article{strath2013guide,
  title={{Guide to the Assessment of Physical Activity: Clinical and Research Applications: A Scientific Statement From the American Heart Association}},
  author={Strath, Scott J and Kaminsky, Leonard A and Ainsworth, Barbara E and Ekelund, Ulf and Freedson, Patty S and Gary, Rebecca A and Richardson, Caroline R and Smith, Derek T and Swartz, Ann M},
  journal={Circulation},
  volume={128},
  number={20},
  pages={2259--2279},
  year={2013},
  publisher={Am Heart Assoc}
}

@book{green1993nonparametric,
  title={{Nonparametric Regression and Generalized Linear Models: A Roughness Penalty Approach}},
  author={Green, Peter J and Silverman, Bernard W},
  year={1993},
  publisher={London: Chapman and Hallz}
}

@article{lin2017locally,
  title={{Locally Sparse Estimator for Functional Linear Regression Models}},
  author={Lin, Zhenhua and Cao, Jiguo and Wang, Liangliang and Wang, Haonan},
  journal={Journal of Computational and Graphical Statistics},
  volume={26},
  number={2},
  pages={306--318},
  year={2017},
  publisher={Taylor \& Francis}
}

@article{cardot2003spline,
  title={{Spline Estimators for the Functional Linear Model}},
  author={Cardot, Herv{\'e} and Ferraty, Fr{\'e}d{\'e}ric and Sarda, Pascal},
  journal={Statistica Sinica},
  volume={13},
  number={3},
  pages={571--591},
  year={2003},
  publisher={JSTOR}
}

@article{fan2001variable,
  title={{Variable Selection Via Nonconcave Penalized Likelihood and Its Oracle Properties}},
  author={Fan, Jianqing and Li, Runze},
  journal={Journal of the American Statistical Association},
  volume={96},
  number={456},
  pages={1348--1360},
  year={2001},
  publisher={Taylor \& Francis}
}

@article{albalak2023setting,
  title={{Setting Your Clock: Associations Between Timing of Objective Physical Activity and Cardiovascular Disease Risk in the General Population}},
  author={Albalak, Gali and Stijntjes, Marjon and van Bodegom, David and Jukema, J Wouter and Atsma, Douwe E and van Heemst, Diana and Noordam, Raymond},
  journal={European Journal of Preventive Cardiology},
  volume={30},
  number={3},
  pages={232--240},
  year={2023},
  publisher={Oxford University Press US}
}

@article{karas2022comparison,
  title={{Comparison of Accelerometry-based Measures of Physical Activity: Retrospective Observational Data Analysis Study}},
  author={Karas, Marta and Muschelli, John and Leroux, Andrew and Urbanek, Jacek K and Wanigatunga, Amal A and Bai, Jiawei and Crainiceanu, Ciprian M and Schrack, Jennifer A},
  journal={JMIR mHealth and uHealth},
  volume={10},
  number={7},
  pages={e38077},
  year={2022},
  publisher={JMIR Publications Toronto, Canada}
}

@article{iso2022physical,
  title={Physical activity as a protective factor for dementia and Alzheimer’s disease: systematic review, meta-analysis and quality assessment of cohort and case--control studies},
  author={Iso-Markku, Paula and Kujala, Urho M and Knittle, Keegan and Polet, Juho and Vuoksimaa, Eero and Waller, Katja},
  journal={British journal of sports medicine},
  volume={56},
  number={12},
  pages={701--709},
  year={2022},
  publisher={BMJ Publishing Group Ltd and British Association of Sport and Exercise Medicine}
}

@article{del2022association,
  title={Association of daily step count and intensity with incident dementia in 78 430 adults living in the UK},
  author={del Pozo Cruz, Borja and Ahmadi, Matthew and Naismith, Sharon L and Stamatakis, Emmanuel},
  journal={JAMA neurology},
  volume={79},
  number={10},
  pages={1059--1063},
  year={2022},
  publisher={American Medical Association}
}

@article{lysen2020actigraphy,
  title={Actigraphy-estimated sleep and 24-hour activity rhythms and the risk of dementia},
  author={Lysen, Thom S and Luik, Annemarie I and Ikram, M Kamran and Tiemeier, Henning and Ikram, M Arfan},
  journal={Alzheimer's \& Dementia},
  volume={16},
  number={9},
  pages={1259--1267},
  year={2020},
  publisher={Wiley Online Library}
}

@article{lim2013sleep,
  title={Sleep fragmentation and the risk of incident Alzheimer's disease and cognitive decline in older persons},
  author={Lim, Andrew SP and Kowgier, Matthew and Yu, Lei and Buchman, Aron S and Bennett, David A},
  journal={Sleep},
  volume={36},
  number={7},
  pages={1027--1032},
  year={2013},
  publisher={Oxford University Press}
}

@article{ning2025accelerometer,
  title={Accelerometer-measured physical activity timing with incident dementia},
  author={Ning, Yuye and Chen, Meilin and Yang, Hao and Jia, Jianping},
  journal={Alzheimer's \& Dementia},
  volume={21},
  number={2},
  pages={e14452},
  year={2025},
  publisher={Wiley Online Library}
}

@article{zhang2023effect,
  title={Effect of physical activity on risk of Alzheimer's disease: a systematic review and meta-analysis of twenty-nine prospective cohort studies},
  author={Zhang, Xiaoqian and Li, Qu and Cong, Wenqiang and Mu, Siyu and Zhan, Rui and Zhong, Shanshan and Zhao, Mei and Zhao, Chuansheng and Kang, Kexin and Zhou, Zhike},
  journal={Ageing Research Reviews},
  volume={92},
  pages={102127},
  year={2023},
  publisher={Elsevier}
}

@article{wiegmann2020alcohol,
  title={Alcohol and dementia--what is the link? A systematic review},
  author={Wiegmann, Caspar and Mick, Inge and Brandl, Eva J and Heinz, Andreas and Gutwinski, Stefan},
  journal={Neuropsychiatric disease and treatment},
  pages={87--99},
  year={2020},
  publisher={Taylor \& Francis}
}

@article{gulpers2016anxiety,
  title={Anxiety as a predictor for cognitive decline and dementia: a systematic review and meta-analysis},
  author={Gulpers, Bernice and Ramakers, Inez and Hamel, Renske and K{\"o}hler, Sebastian and Oude Voshaar, Richard and Verhey, Frans},
  journal={The American Journal of Geriatric Psychiatry},
  volume={24},
  number={10},
  pages={823--842},
  year={2016},
  publisher={Elsevier}
}

@article{duong2017dementia,
  title={Dementia: What pharmacists need to know},
  author={Duong, Silvia and Patel, Tejal and Chang, Feng},
  journal={Canadian Pharmacists Journal/Revue des Pharmaciens du Canada},
  volume={150},
  number={2},
  pages={118--129},
  year={2017},
  publisher={SAGE Publications Sage CA: Los Angeles, CA}
}

@article{wilkinson2019identifying,
  title={Identifying dementia outcomes in UK Biobank: a validation study of primary care, hospital admissions and mortality data},
  author={Wilkinson, Tim and Schnier, Christian and Bush, Kathryn and Rannikm{\"a}e, Kristiina and Henshall, David E and Lerpiniere, Chris and Allen, Naomi E and Flaig, Robin and Russ, Tom C and Bathgate, Deborah and others},
  journal={European journal of epidemiology},
  volume={34},
  number={6},
  pages={557},
  year={2019}
}

@article{raichlen2023sedentary,
  title={Sedentary behavior and incident dementia among older adults},
  author={Raichlen, David A and Aslan, Daniel H and Sayre, M Katherine and Bharadwaj, Pradyumna K and Ally, Madeline and Maltagliati, Silvio and Lai, Mark HC and Wilcox, Rand R and Klimentidis, Yann C and Alexander, Gene E},
  journal={Jama},
  volume={330},
  number={10},
  pages={934--940},
  year={2023},
  publisher={American Medical Association}
}

@article{meng2025insomnia,
  title={Insomnia and risk of all-cause dementia: A systematic review and meta-analysis},
  author={Meng, Mingxian and Shen, Xiaoming and Xie, Yanming and Lan, Rui and Zhu, Shirui},
  journal={PloS one},
  volume={20},
  number={4},
  pages={e0318814},
  year={2025},
  publisher={Public Library of Science San Francisco, CA USA}
}

@article{boyd2011distributed,
  title={Distributed optimization and statistical learning via the alternating direction method of multipliers},
  author={Boyd, Stephen and Parikh, Neal and Chu, Eric and Peleato, Borja and Eckstein, Jonathan and others},
  journal={Foundations and Trends{\textregistered} in Machine learning},
  volume={3},
  number={1},
  pages={1--122},
  year={2011},
  publisher={Now Publishers, Inc.}
}

@article{zhou2025heterogeneous,
  title={Heterogeneous functional regression for subgroup analysis},
  author={Zhou, Yeqing and Jiang, Fei},
  journal={Journal of Computational and Graphical Statistics},
  volume={34},
  number={3},
  pages={872--883},
  year={2025},
  publisher={Taylor \& Francis}
}
\endgroup


\end{document}